# Terphenylthiazole-based self-assembled monolayers on cobalt with high conductance photo-switching ratio for spintronics

*Vladimir Prudkovskiy,[a,#] Imane Arbouch,[b,#] Anne Léaustic,[c] Pei Yu,[c,]\**
*Colin Van Dyck,[b] David Guérin,[a] Stéphane Lenfant,[a]*
*Talal Mallah,[c] Jérôme Cornil,[b,]\* and Dominique Vuillaume.[a,]\**

a) Institute for Electronics Microelectronics and Nanotechnology (IEMN), CNRS, Av. Poincaré, Villeneuve d'Ascq, France.

b) Laboratory for Chemistry of Novel Materials, University of Mons, Place du Parc 20, 7000 Mons, Belgium.

c) Institut de Chimie Moléculaire et des Matériaux d'Orsay (ICMMO), CNRS, Université Paris-Saclay, 91405 Orsay Cedex, France

pei.yu@universite-paris-saclay.fr
jerome.cornil@umons.ac.be
dominique.vuillaume@iemn.fr

Two new photo-switchable terphenylthiazoles molecules are synthesized and self-assembled as monolayers on Au and on ferromagnetic Co electrodes. The electron transport properties probed by conductive atomic force microscopy in ultra-high vacuum reveal a conductance of the light-induced closed (c) form larger than for the open (o) form. We report an unprecedented conductance ratio up to 380 between the closed and open forms on Co for the molecule with the anchoring group (thiol) on the side of the two N atoms of the thiazole unit. This result is rationalized by Density Functional Theory (DFT) calculations coupled to the Non-Equilibrium Green's function (NEGF) formalism. These calculations show that the high conductance in the closed form is due to a strong electronic coupling between the terphenylthiazole molecules and the Co electrode that manifests by a resonant transmission peak at the Fermi energy of the Co electrode with a large broadening. This behavior is not observed for the same molecules self-assembled on gold electrodes. These high conductance ratios make these Co-based molecular junctions attractive candidates to develop and study switchable molecular spintronic devices.

# INTRODUCTION.

The use of photochromic and electrochromic active molecules in molecular junctions allows modulating the electron- and magneto-transport response of these junctions using light or electric field.[1-5] When these molecules are assembled on ferromagnetic (FM) electrodes, the spin-polarized electron transport through the FM/molecules/FM junctions is expected to depend on the conformation of the molecules and on the molecule/electrode atomic contact geometry, as evaluated from theoretical studies.[6-8] For instance, a huge increase of the magnetoresistance (MR) ratio was calculated for an azobenzene-based molecular junction (from 65% for the *cis* conformation to 2700% for the *trans* conformation);[8] moreover, the voltage dependence of the MR is significantly different for a dithienylethene molecule in its *open* and *closed* forms.[7] These studies suggest new approaches to control the spin-polarized electron transport at the molecule-ferromagnetic hybrid interfaces. The building of new device functionalities beyond conventional spintronics by designing active metal/ functional molecule combinations is highly desirable to exploit this unique tailoring opportunity offered by chemistry. However, the experimental demonstration of such electro-optical molecular spintronic devices is lacking.

We have recently reported the optically induced conductance switching at the nanoscale (via conductive-AFM measurements) of diarylethene derivatives in self-assembled monolayers (SAMs) on $La_{0.7}Sr_{0.3}MnO_3$ electrodes;[9] we observed a weak current (I) switching of the diarylethene molecular junctions (*closed* form/ *open* form conductance ratio I(closed)/I(open)=$R_{c/o}$ <8), partly hidden under some conditions by the optically induced conductance switching of the $La_{0.7}Sr_{0.3}MnO_3$ substrate. We have measured a slightly higher conductance ratio (between the *cis/trans* isomers) of about 20 for azobenzene derivatives on Co.[10] These performances are lower than those of the same self-assembled monolayers on gold electrodes for which conductance ratios up to ≈7x10$^3$ were measured for azobenzene derivatives[4] and $R_{c/o}$ ≈ 100 were calculated and measured for diarylethene derivatives.[11, 12]

Here, we report an unprecedented $R_{c/o}$ up to ≈ 380 for terphenylthiazole-based SAMs on Co. Two thiazole-based terarylene molecules, (scheme 1, referred to as TPT(A) and TPT(B) in the following) were designed, synthesized and the electron transport properties of SAMs on Au and Co were measured by conductive atomic force microscopy (CAFM in air and UHV, respectively). At a theoretical level, we carried out first principle calculations on these molecular junctions by means of the state-of-the art Non-Equilibrium Green's Function method (NEGF) combined with Density Functional Theory (DFT).[13] The transmission spectra at different biases are calculated for the various TPT(B)-based molecular junctions and I-V curves simulated using the Landauer Büttiker formalism within the coherent transport regime.[14] The experimental and theoretical data are consistent and lead to the main conclusions that: (i) TPT(B) based junctions exhibit higher $R_{c/o}$ compared to TPT(A) junctions on both Au and Co (by a factor ≈ 2-2.5); (ii) the TPT SAMs on Co yield higher $R_{c/o}$ than their counterparts on Au (by a factor ≈ 20-25) and (iii) TPT(B) SAMs on Co exhibit the highest $R_{c/o}$ ratio (≈195-380 from experiments, ≈110-290 from theory). The latter finding is rationalized by a strong coupling between the closed form of the TPT(B) molecule and the Co electrode manifested by a resonant transmission peak with a large broadening at the Fermi level of the Co electrode.

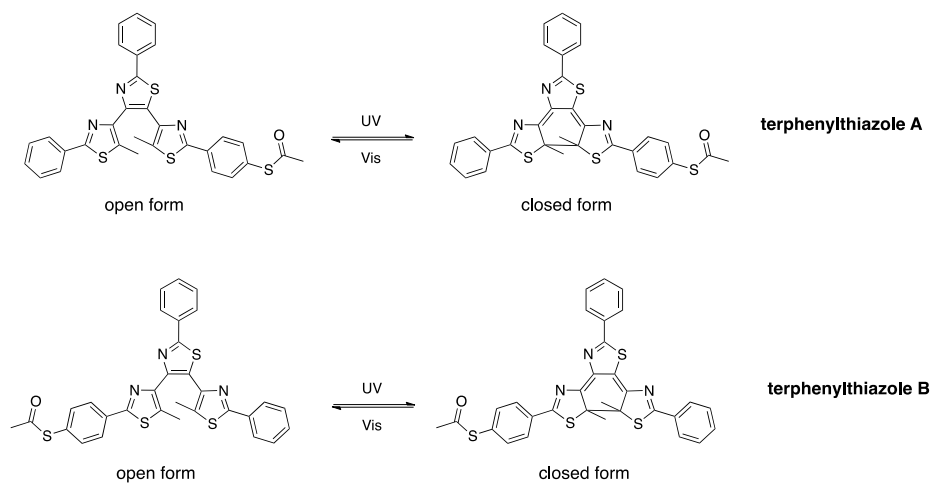

**Scheme 1**. *Photochemical transformation of terphenylthiazole molecules TPT( A) and TPT(B).*

## Experimental Results

We have chosen to work with two thiol functionalized thiazole-based terarylenes (molecules TPT(A) and TPT(B)) that share the same photochromic core structure, but differ with the position of the anchoring group with respect to the central thiazole (Scheme 1). Terarylenes are known to possess generally good and widely tunable photochromic properties.[15-19] Moreover, unlike classic diarylethenes with standard ethene bridges (perfluorocyclopentene, maleimide etc.),[1, 20] terarylenes offer additional flexibility in terms of functionalization with the central aryl site.[21-23] The choice of thiazole as aryl over the much more documented thiophene is motivated by the fact that thiazole-based terarylenes are known to be thermally more stable in their closed form than their thiophene-based counterparts.[15] Finally, the use of thiazole (or thiophene) as the central ethene moiety introduces a dissymmetry within the molecular switch, absent for classic diarylethenes. Even though the optical properties of the two molecules are expected to be quite similar both in their open and closed forms in the isolated state, it is interesting to investigate how such a subtle dissymmetry could affect the transport properties in junctions. The detailed syntheses of the two new thiazole-based photoswitches and their full characterizations using standard techniques can be found in the Supporting Information (section 1). We checked the photo-switching of the two molecules in solution (µM in $CH_2Cl_2$) by UV-vis spectroscopy. The two molecules display the expected photochromic properties and can be reversibly and cleanly switched between a colorless open form and a deeply colored closed form, with very similar spectral evolutions (see in the Supporting Information).

We fabricated the SAMs on freshly prepared template-stripped $^{TS}$Au and evaporated Co surfaces from diluted millimolar solutions of TPT(A) and TPT(B) in ethanol/tetrahydrofuran (EtOH/THF 80:20 v/v) (see methods and section 2 in the Supporting Information). The thicknesses of the SAMs measured by ellipsometry are 9-10 ± 2 Å for TPT(A) and TPT(B) on $^{TS}$Au. Given the theoretical estimated length of the molecule of 16.3 Å (in agreement with X-ray crystallography data)[15] we estimate a molecular tilt angle of ≈ 54 ± 9° in these SAMs. On Co substrate, the thickness measurements are less reliable because of the slightly higher roughness of the Co surface (≈0.8 nm[24] vs. ≈0.4 nm for our $^{TS}$Au surface[25, 26]) and the need to use a specific sealed cell under dry $N_2$ (see methods and section 4 in the Supporting Information). We grossly estimate a thickness between 6 and 17 Å for TPT(A) SAM (i.e. 11.5 ± 5.5 Å) and 12 -22 Å (17 ± 5 Å) for the TPT(B) SAMs.

The analysis of the X-ray Photoelectron Spectroscopy (XPS) data of SAMs of TPT(A) and TPT(B) on $^{TS}$Au shows the presence of C, N and S in the SAMs. The C 1s major peak at 284.8 eV (section 5 in the Supporting Information) corresponds to C-C, C-N and C-S bonds. The shoulder observed at 286.2 eV is assigned to the three S-C=N carbons of the thiazole rings.[27] The S 2p region shows two doublets (S $2p_{1/2}$ and S $2p_{3/2}$) associated to the S-C and S-Au bonds (section 5 in the Supporting Information) and the amplitude ratios [S-Au]/[S-C] are ≈ 0.4-0.5, slightly higher than the expected ratio of 1/3. For both molecules, in addition to the main N 1s peak associated to the C=N bond, a second peak is also observed and is likely associated to a "coordinated-like" form (C=N**...**Au) of nitrogen (section 5 in the Supporting Information). This N 1s peak splitting is observed when the N atoms interact with a metal surface,[28-30] a possible situation considering the large tilt angle of the molecules inferred from thickness measurements and calculations (geometry optimization, *vide infra*). This "coordinated-like" peak is further used as a fingerprint of a strong interaction between the molecule and the substrate in correlation with the electron transport properties of the molecular junctions (*vide infra*). Interestingly, the ratio of the peak amplitude [C=N**...**Au]/[N=C] is higher for TPT(B) than for TPT(A): 0.74 vs. 0.34, respectively (Fig. S5 and Table S1 in the Supporting information). This feature is likely related with the asymmetry of the two molecules (*vide infra*, section theory and discussion). The XPS spectra of the SAMs of TPT(A) and TPT(B) on Co show the presence of C, N and S elements. As for the molecules on $^{TS}$Au, the C 1s peak is composed of C-C, C-N and C-S signals (section 5 in the Supporting Information). In the S 2p region (section 5 in the Supporting Information), we observe the contribution of S-C and S-Co bonds and the amplitude ratio [S-Co]/[S-C] ≈ 0.5-0.6 is higher than the expected value of 1/3. This feature again suggests that more than one S atom is bound to the Co, a situation that cannot be excluded because of the large tilt angle of the molecule that is consistent with the geometry optimization calculations (*vide infra*). The N 1s region (section 5 in the Supporting Information) also shows the two peaks of the C=N bonds and the [C=N**...**Co] one but the ratio of the amplitudes is different for the TPT(B), [C=N**...**Co]/[N=C]=1.24, and TPT(A) molecules, [C=N**...**Co]/[N=C]=0.39. As for the SAMs on Au, this is most likely attributed due to the interaction of N atoms with the surface resulting from the large molecule tilt. Both on Au and Co, these ratios thus point out to a higher concentration of N atoms interacting with the surface for TPT(B) than TPT(A), highlighting a difference in the interaction of the two isomers with the substrates. Finally, the O 1s region reveals (Fig. S6 in the Supporting Information) a residual oxidized Co[31] as in our previous work on azobenzene derivatives on Co (Fig. S3 in Ref. 10), albeit the protocol and precautions used during the grafting and measurements.

Figure 1 shows the dataset of measured current-voltage (I-V) curves of the SAMs of TPT(A) on Co, measured by CAFM in UHV for (a) the pristine SAM (as fabricated with open form), (b) after UV light illumination and (c) after white light illumination (see methods and Supporting Information: section 6 for CAFM and section 7 for illumination conditions). In the presented dataset, we removed the I-V traces reaching the sensitivity limit of the apparatus (almost flat I-V traces and/or displaying random staircase behavior in the range of 0.1 to few pA), those reaching the saturating current of the preamplifier during the voltage scan and those with large and abrupt changes in the measured current (CAFM tip contact issue) - see section 6 in the Supporting Information (Figs. S7 and S8). The datasets are shown as 2D-histograms (heat maps) and the red line is the calculated mean $\bar{I}$-V curve (the same datasets plotted in regular style, not as a heat map, are also given in section 6 of the Supporting Information; Figs. S9). We observe that the TPT(A) molecules have a higher conductance in the closed form (TPT(A)-c) after UV-light exposure than in the open (TPT(A)-o) form (Figs. 1a and 1b), as already observed for other diarylethene derivatives.[1, 2, 11, 12] The histograms of the currents taken at 0.5V (in blue) and -0.5V (in red) are shown in Figs 1d-1f. They are fitted by a log-normal distribution. Table 1 summarizes the fitted values of the log-mean current (log-$\bar{I}$) and the log-standard deviation (log-σ) for the pristine molecular junctions and after UV and visible light irradiation and Fig. 5 shows the evolution of the mean current. We calculate the mean on/off (closed/open) current ratios $R_{c/o}$(TPT(A)-Co) ≈ 150-180 at both 0.5V and -0.5V as the ratio of the mean current values of the histograms after UV irradiation and of the pristine SAMs. We also calculated the $R_{c/o}$-V curve from the ratio of the corresponding mean $\bar{I}$-V curves in Figs. 1a and 1b ($R_{c/o}(V)=\bar{I}_{UV}(V)/\bar{I}_{pristine}(V)$) and found values of the same order of magnitude (107-214) over the whole voltage range (Fig. S10, section 6 in the Supporting information). Figure 2 shows the dataset for the TPT(B) molecules on Co. We observe the same trends as for the TPT(A) molecules, although the closed/open current ratio is slightly larger, $R_{c/o}$(TPT(A)-Co) ≈ 320 and 250 at 0.5V and -0.5V (and 195-380 for the entire voltage range, Fig. S9 section 6 in the Supporting Information) considering the HC (high current) peak for the closed form in the measured current distribution (Fig. 2e). We note, however, two differences for the SAMs of TPT(B) on Co : (i) we observe a larger current dispersion; and (ii) the open to close transformation is not complete. The current histogram in Fig. 2e shows that a fraction (about 50 % of the counts, as estimated from the calculated areas under the LC and HC peaks) is reminiscent of the open state in view of a second peak (low current peak, LC peak) centered in the 0.1-1 nA range as for the pristine SAM and after white light exposure. The large dispersion of current and the presence of two peaks in some cases can also be due to disorder induced by a slightly higher roughness of the Co electrode compared to Au as demonstrated by a series of electron transport measurements on electrodes with various controlled topography.[32] Finally, we note that TPT(B) molecules have a larger c/o current ratio than TPT(A) molecules, in agreement with the theoretical results (*vide infra*). We also note that the open to close switching is reversible under white light illumination for the two molecular junctions (we have not done a detailed cyclability study, but the molecules were photoswitched open/closed twice). The lack of a complete photizomerization for TPT(B) on Co may be rationalized by the increase in accessible excited states originating from the strong coupling to the Co metal surface (*vide infra*, theory section).[33]

For comparison purpose, we also characterize by CAFM (in air) the electron transport properties of the TPT SAMs on $^{TS}$Au substrates (see methods and section 6 in the Supporting Information). Figures 3 and 4 show the 2D I-V histograms of the molecular junctions of TPT(A) and TPT(B) on an $^{TS}$Au substrate (methods and section 6 in the Supporting Information). As for the SAMs on Co, we still observe that the conductance of the SAM is higher with the molecules in the closed form than in the open form. The main difference is that the closed/open current ratios are lower for the SAMs on gold than on Co, for both molecules. We obtain $R_{c/o}$(TPT(A)-Au) ≈ 7 at 0.5 V and -0.5 V (Fig. 3) (and the same value, 6-8, in the whole voltage range - Fig. S9, section 6 Supporting Information) and a current ratio $R_{c/o}$(TPT(B)-Au) ≈ 16 at 0.5 V and 14 at -0.5 V for TPT(B) with a value between 9 and 18 for the whole voltage range (Fig. S10, section 6 in the Supporting Information). We still measure a slightly higher closed/open conductance ratio for the TPT(B) than for TPT(A) molecules (as for the SAMs on Co). Figure 5 and Table 1 summarize the evolution of the mean current $\bar{I}$ for the four molecular junctions.

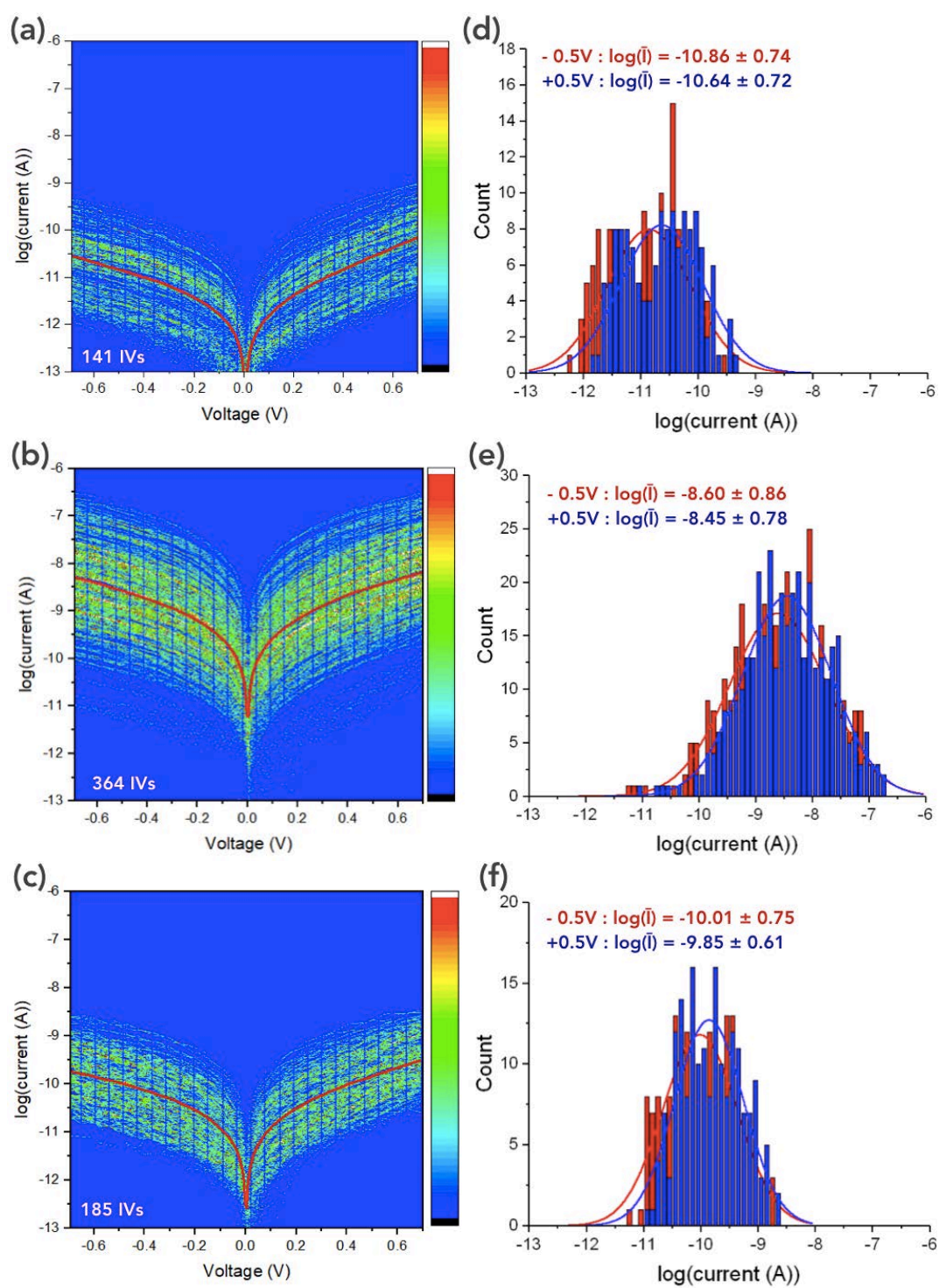

*Figure 1.* 2D histograms of the current-voltage (I-V) curves: (a) pristine SAM of TPT(A) on Co, (b) after UV irradiation, (c) after visible light irradiation. The *currents are measured by CAFM in UHV. The number of I-V traces in the dataset is shown on the figures. The red line is the mean $\bar{I}$ current. Histograms of the currents at 0.5V and -0.5V for (d) pristine SAM of TPT(A), (e) after UV irradiation, (f) after visible light irradiation. The fit parameters of the log-normal distribution, log-$\bar{I}$ (log-mean current) and log-σ (log-standard deviation), are given in the figures and summarized in Table 1.*

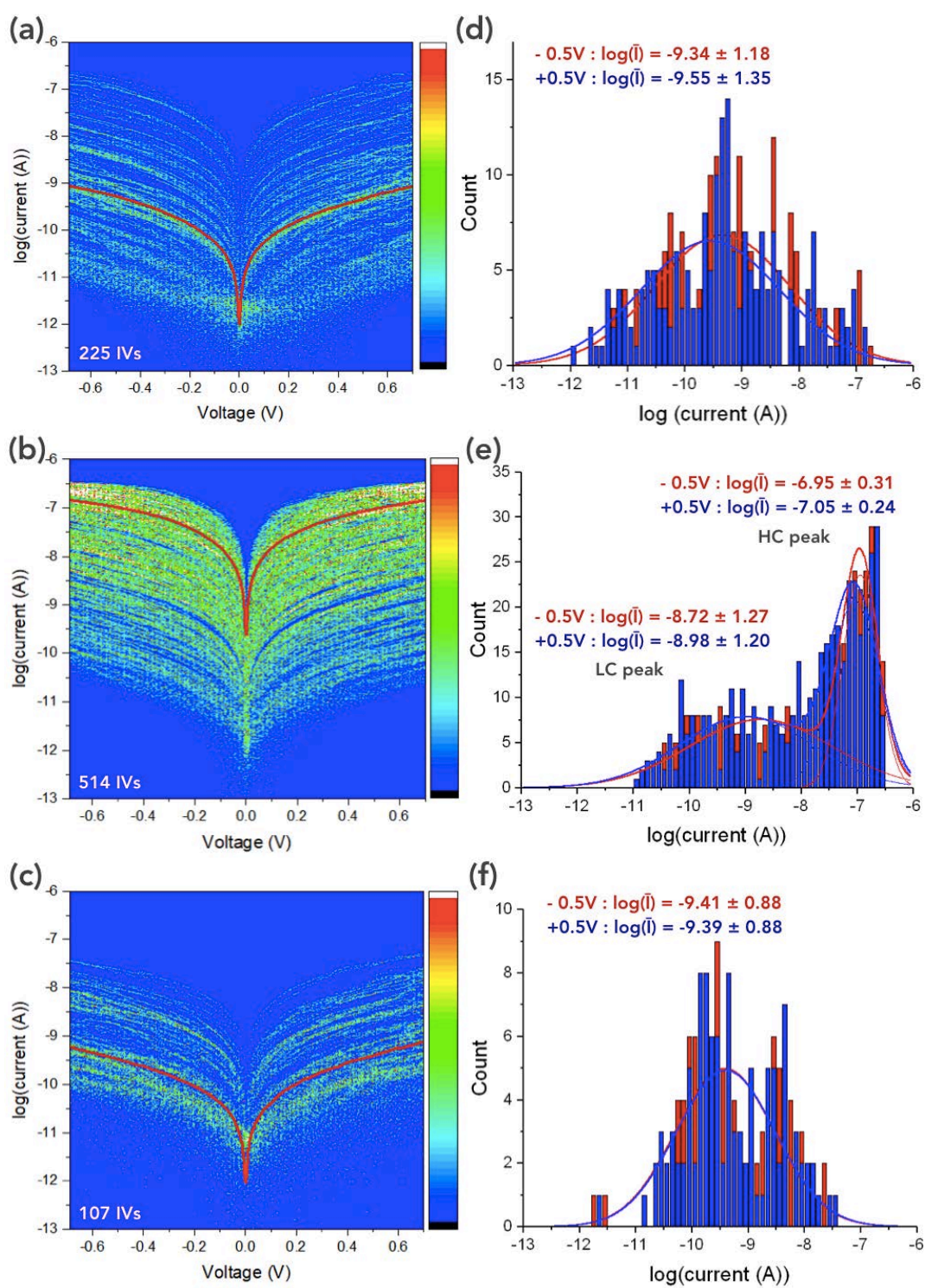

*Figure 2*. 2D histograms of the current-voltage (I-V) curves: (a) pristine SAM of TPT(B) on Co, (b) after UV irradiation, (c) after visible light irradiation. The *currents are measured by CAFM in UHV. The number of I-V traces in the dataset is shown on the figures. The red line is the mean $\bar{I}$ current. Histograms of the currents at 0.5V and -0.5V for (d) pristine SAM of TPT(B), (e) after UV irradiation, (f) after visible light irradiation. The fit parameters of the log-normal distribution, log-$\bar{I}$ (log-mean current) and log-σ (log-standard deviation), are given in the figures and summarized in Table 1.*

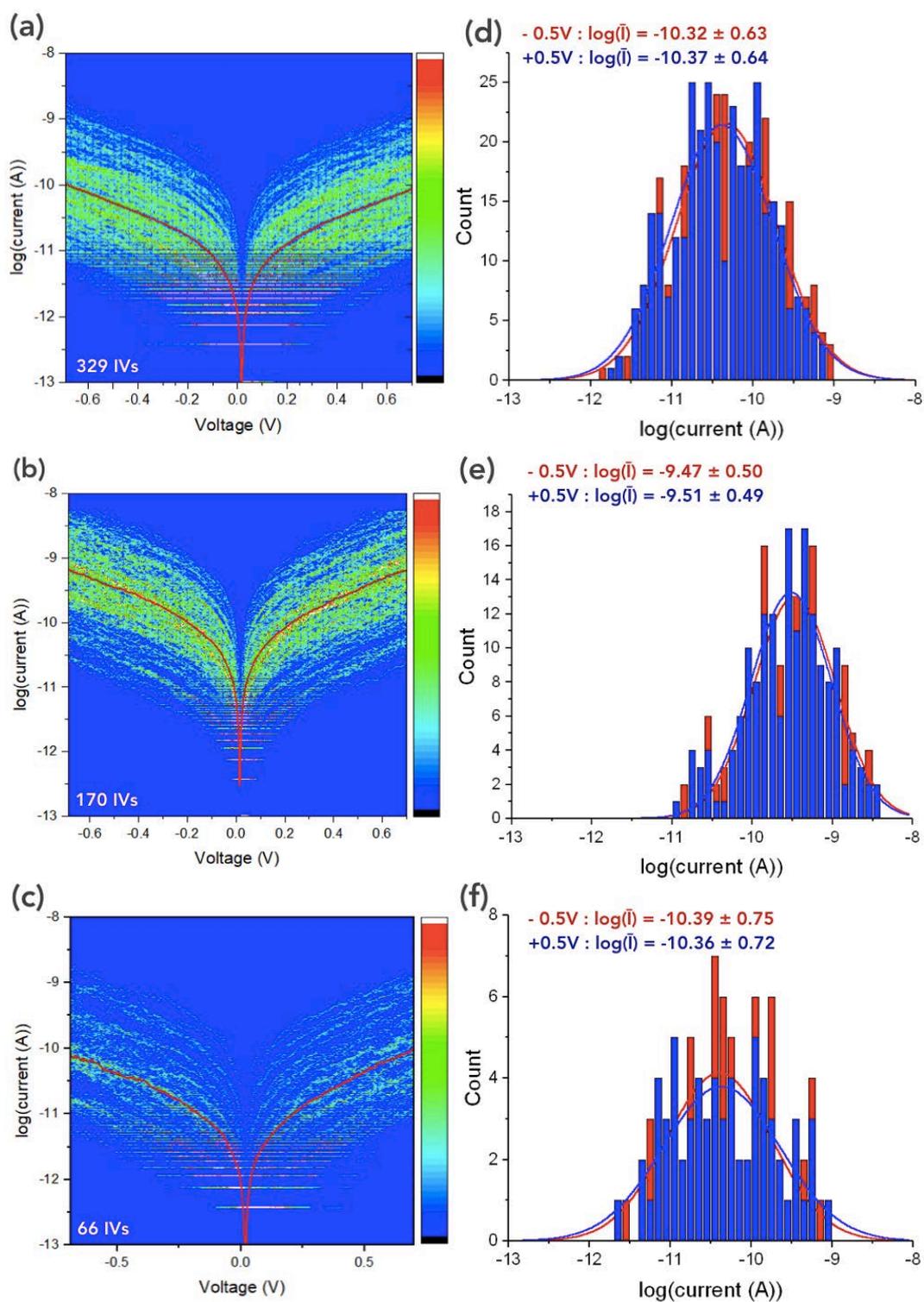

*Figure 3*. 2D histograms of the current-voltage (I-V) curves: (a) pristine SAM of TPT(A) on ᵀˢAu, (b) after UV irradiation, (c) after visible light irradiation. The *currents are measured by CAFM in air. The number of I-V traces in the dataset is shown on the figures. The red line is the mean $\bar{I}$ current. Histograms of the currents at 0.5V and -0.5V for (d) pristine SAM of TPT(A), (e) after UV irradiation, (f) after visible light irradiation. The fit parameters of the log-normal distribution, log-$\bar{I}$ (log-mean current) and log-σ (log-standard deviation), are given in the figures and summarized in Table 1.*

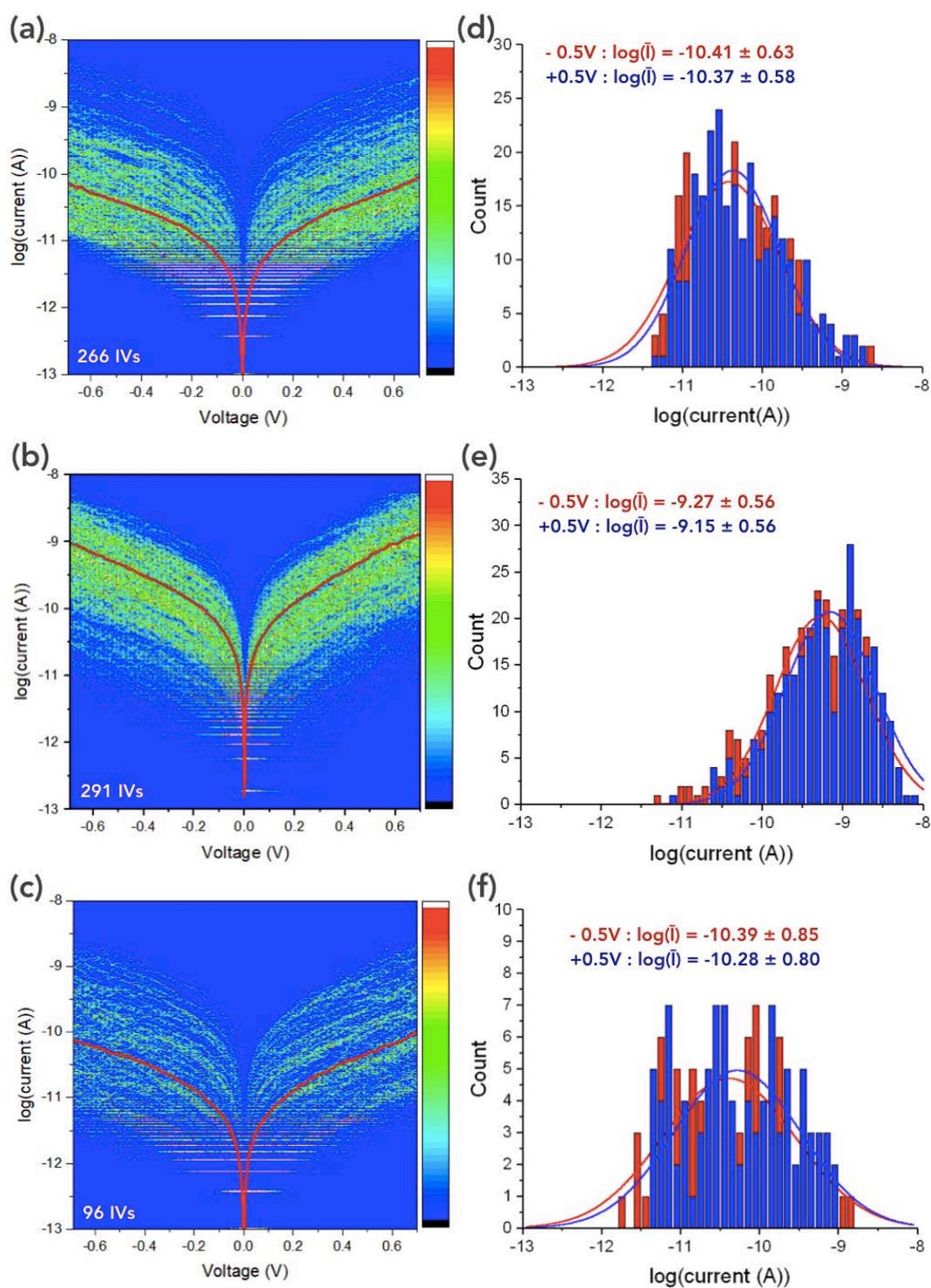

*Figure 4*. 2D histograms of the current-voltage (I-V) curves: (a) pristine SAM of TPT(B) on $^{TS}$Au, (b) after UV irradiation, (c) after visible light irradiation. The *currents are measured by CAFM in air. The number of I-V traces in the dataset is shown on the figures. The red line is the mean $\bar{I}$ current. Histograms of the currents at 0.5V and -0.5V for (d) pristine SAM of TPT(B), (e) after UV irradiation, (f) after visible light irradiation. The fit parameters of the log-normal distribution, log-$\bar{I}$ (log-mean current) and log-σ (log-standard deviation), are given in the figures and summarized in Table 1.*

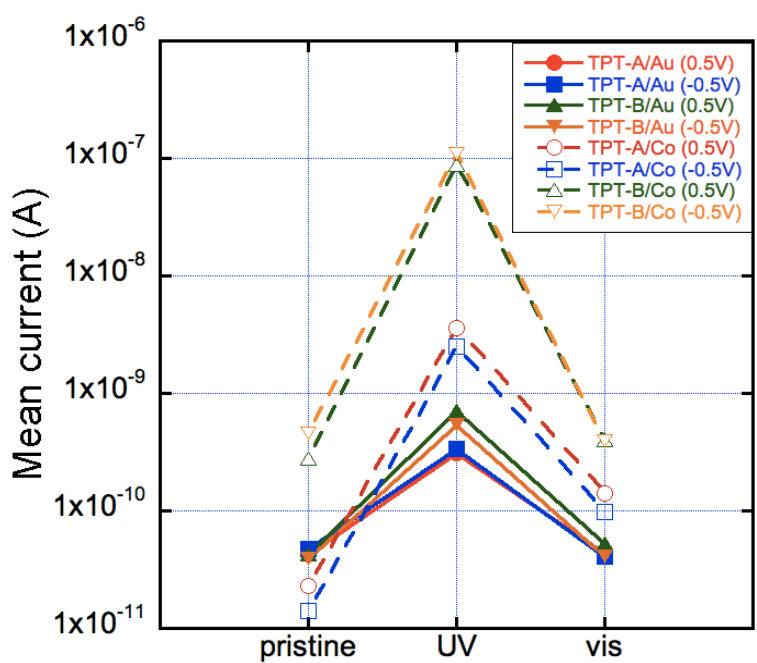

***Figure 5.*** *Evolution of the mean current Ī (deduced from the log-normal fits of the current distributions in Figs. 1-4) at 0.5 V and -0.5 V for the four molecular junctions (see inset) and under the three conditions: pristine, after UV and visible light illumination.*

|  | TPT(A)-Au | | | TPT(B)-Au | | |
|---|---|---|---|---|---|---|
|  | pristine | UV | vis | pristine | UV | vis |
| log-Ī ± log-σ at 0.5V at -0.5V | -10.37 ± 0.64<br>-10.32 ± 0.63 | -9.51 ± 0.49<br>-9.47 ± 0.50 | -10.36 ± 0.72<br>-10.39 ± 0.75 | -10.37 ± 0.58<br>-10.41 ± 0.63 | -9.15 ± 0.56<br>-9.27 ± 0.56 | -10.28 ± 0.80<br>-10.39 ± 0.85 |
| Ī(A) at 0.5V at -0.5V | $4.3 \times 10^{-11}$<br>$4.8 \times 10^{-11}$ | $3.1 \times 10^{-10}$<br>$3.4 \times 10^{-10}$ | $4.4 \times 10^{-11}$<br>$4.1 \times 10^{-11}$ | $4.3 \times 10^{-11}$<br>$3.9 \times 10^{-11}$ | $7.1 \times 10^{-10}$<br>$5.4 \times 10^{-10}$ | $5.3 \times 10^{-11}$<br>$4.1 \times 10^{-11}$ |
| $R_{c/o}$ at 0.5V at -0.5V |  | 7.2<br>7.1 |  |  | 16.6<br>13.8 |  |
|  | TPT(A)-Co | | | TPT(B)-Co | | |
|  | pristine | UV | vis | pristine | UV | vis |
| log-Ī ± log-σ at 0.5V at -0.5V | -10.64 ± 0.72<br>-10.86 ± 0.74 | -8.45 ± 0.78<br>-8.60 ± 0.86 | -9.85 ± 0.61<br>-10.01 ± 0.75 | -9.55 ±1.35<br>-9.34 ± 1.18 | HC peak<br>-7.05 ± 0.24<br>-6.95 ± 0.31<br>LC peak<br>-8.98 ±1.20<br>-8.72 ± 1.27 | -9.39 ± 0.88<br>-9.41 ± 0.88 |
| Ī(A) at 0.5V at -0.5V | $2.3 \times 10^{-11}$<br>$1.4 \times 10^{-11}$ | $3.6 \times 10^{-9}$<br>$2.5 \times 10^{-9}$ | $1.4 \times 10^{-10}$<br>$9.8 \times 10^{-11}$ | $2.8 \times 10^{-10}$<br>$4.6 \times 10^{-10}$ | HC peak<br>$8.9 \times 10^{-8}$<br>$1.1 \times 10^{-7}$<br>LC peak<br>$1.1 \times 10^{-9}$<br>$1.9 \times 10^{-9}$ | $4.0 \times 10^{-10}$<br>$3.9 \times 10^{-10}$ |
| $R_{c/o}$ at 0.5V at -0.5V |  | 155<br>182 |  |  | 318<br>245 |  |

*Table 1. Parameter values of the log-normal distributions fitted on the current histograms taken at 0.5 V and -0.5 V : log-mean current (log-Ī), the log-standard deviation (log-σ), mean current Ī and corresponding closed/open current ratios ($R_{c/o}$) calculated as the ratios of the current after UV light irradiation (HC peak in the case of TPT(B)-Co sample) over those for the pristine (open form) samples.*

We analyzed the I-V curves by a simple analytical model: the single energy level (SEL) model, which assumes that the electron transport through the molecular junction is mediated by only one molecular orbital (section 6 in the Supporting Information). Briefly, all the I-V curves in the datasets shown in Figs. 1 to 4 are individually fitted by the usual SEL equation:[34]

$$I = N\frac{8e}{h}\frac{\Gamma_1\Gamma_2}{\Gamma_1+\Gamma_2}\left[\arctan\frac{2\varepsilon_0 + eV\frac{\Gamma_1-\Gamma_2}{\Gamma_1+\Gamma_2} + eV}{2(\Gamma_1+\Gamma_2)} - \arctan\frac{2\varepsilon_0 + eV\frac{\Gamma_1-\Gamma_2}{\Gamma_1+\Gamma_2} - eV}{2(\Gamma_1+\Gamma_2)}\right] \quad (1)$$

with N the number of molecules in the junction, e the electron charge, h the Planck's constant. The fitted parameters are $\varepsilon_0$ the energy position (with respect to the Fermi energy of electrodes) of the molecular orbital involved in the electron transport, $\Gamma_1$ and $\Gamma_2$ the coupling energy between the molecules and the two electron reservoirs (electrodes). We limited the fits to a voltage window -0.5 V to 0.5 V for the best accurate fits (see section 6 in the Supporting Information for details on the fit protocol, Figs. S11 and S12) and also mainly because above |0.5|V the HOMO and LUMO levels of the closed form of TPTs start to contribute to the electron transport (theory section, *vide infra*), a case not consistent with the SEL. This model assumes that the molecular orbital broadening is described by a Lorentzian or Breit-Wigner distribution.[34, 35] This is clearly not the case for the TPT molecules on Co (theory section, *vide infra*) and the SEL model was not used in this case. The exact number of molecules, N, varies with the details of the tip shape, loading force, Young modulus of the SAMs. We have used N = 10 for all the fits (see section 6 in the Supporting Information). The use of N as a multiplication factor in Eq. (1) means that we neglect the intermolecular interactions,[36-38] so that the fitted values of $\Gamma_1$ and $\Gamma_2$ can only be used for a relative/qualitative comparison between the two states of the same junction and not for a quantitative comparison with theory or other experiments such as single-molecule measurements.

Figures 6 and 7 give the distributions of the parameters $\varepsilon_0$, $\Gamma_1$ and $\Gamma_2$ extracted by fitting Eq. 1 on the datasets of TPT(A) and TPT(B) on $^{TS}$Au shown in Figs. 3 and 4, as well as the fit by a Gaussian distribution giving the mean values of these parameters. Typical fits of this model on the mean Ī-V curves for each sample and the three conditions (pristine, UV and visible light exposures) are also given section 6 in the Supporting Information (Fig. S12). In both cases, the energy gap between the resonant and the Fermi levels is lowered upon switching from the open to the closed form of the molecules: from 0.49 eV to 0.4 eV for TPT(A)-o and TPT(A)-c and from 0.52 eV for TPT(B)-o to 0.34 eV for TPT(B)-c. The energy level is lowered by 0.10-0.18 eV with a more important effect for TPT(B) molecules. The increase of the current for the closed form is also due to a significant increase of the molecule-electrode coupling by a factor ≈3.5; the average coupling energies, $\Gamma_1$ and $\Gamma_2$, increase from 0.11-0.20 meV (open form), to 0.46-0.59 meV (closed form) with a larger distribution in this latter case (Figs. 6d-f and Figs. 7d-f) - (the data are summarized in Table 2). Albeit the fact that the electrode-molecule interfaces are different chemically and physically (chemical grafting on flat and large area Au surface on one side, nanoscale mechanical contact with PtIr tip on the other side), the I-Vs are quite symmetrical (and thus the SEL model gives almost the same values of $\Gamma_1$ and $\Gamma_2$). However, it was shown that asymmetrical molecule-electrode couplings do not systematically result in asymmetrical I-Vs.[39, 40]

|  | TPT(A)-Au | | | TPT(B)-Au | | |
|---|---|---|---|---|---|---|
|  | pristine | UV | vis | pristine | UV | vis |
| $\varepsilon_0$ (eV) | 0.49 ± 0.09 | 0.40 ± 0.07 | 0.49 ± 0.09 | 0.52 ± 0.07 | 0.34 ± 0.07 | 0.51 ± 0.09 |
| $\Gamma_1$ (meV) | 0.14 ± 0.24 | 0.50 ± 0.43 | 0.11 ± 0.23 | 0.16 ± 0.08 | 0.46 ± 0.41 | 0.10 ± 0.18 |
| $\Gamma_2$ (meV) | 0.16 ± 0.24 | 0.49 ± 0.40 | 0.17 ± 0.17 | 0.20 ± 0.11 | 0.59 ± 0.43 | 0.18 ± 0.18 |

*Table 2. Values (and their standard variations) for the Gaussian fits to determine the energy position $\varepsilon_0$ (with respect to the Fermi energy of electrodes) of the molecular orbital involved in the electron transport, $\Gamma_1$ and $\Gamma_2$ the coupling energy between the molecules and the two electrodes.*

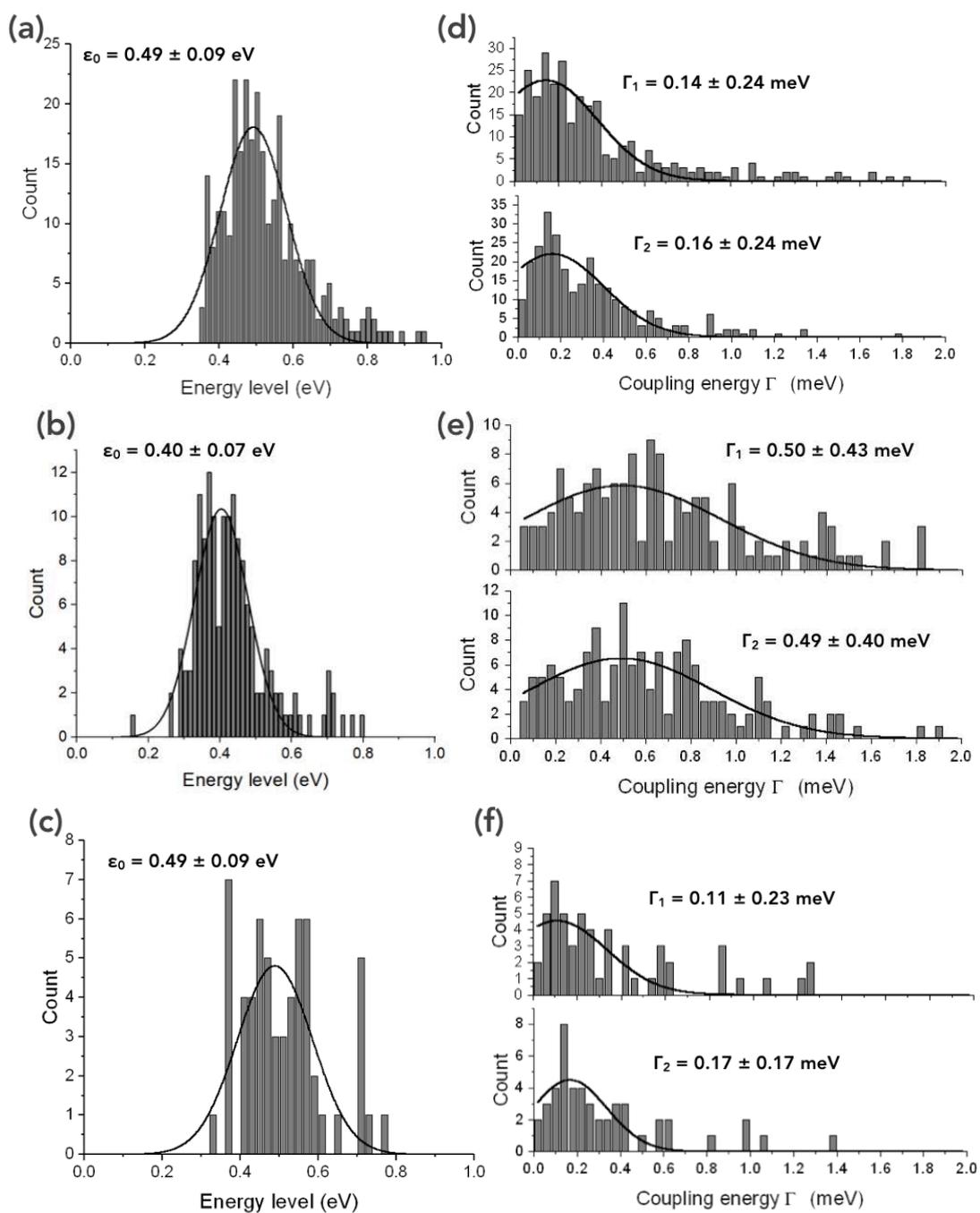

*Figure 6*. *Histograms of the fitted energy $\varepsilon_0$, $\Gamma_1$ and $\Gamma_2$ for the TPT(A) molecules on $^{TS}Au$ : (a,d) pristine sample, (b,e) after UV light illumination and (c,f) after white light illumination.*

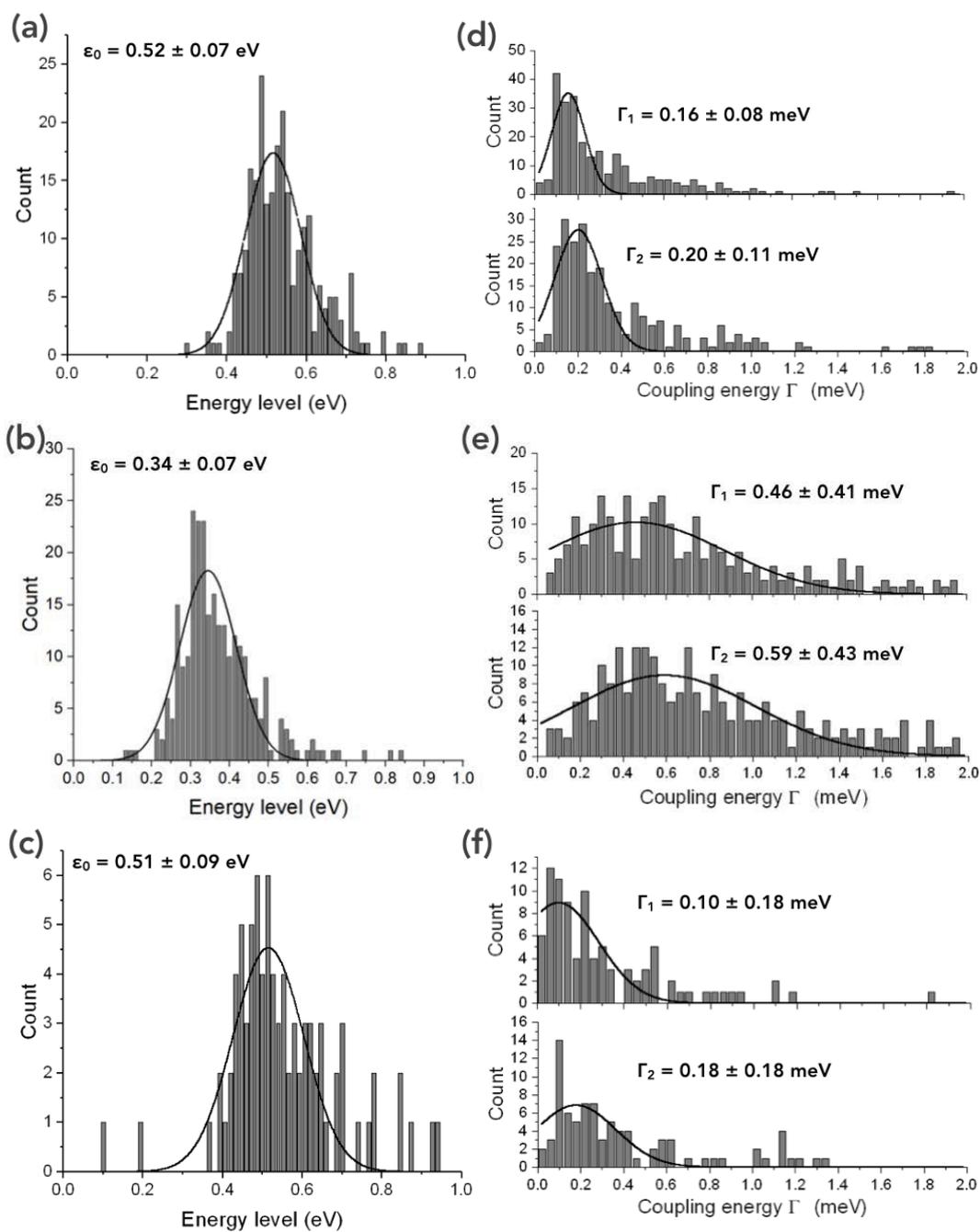

*Figure 7. Histograms of the fitted energy $\varepsilon_0$, $\Gamma_1$ and $\Gamma_2$ for the TPT(B) molecules on $^{TS}Au$ : (a,d) pristine sample, (b,e) after UV light illumination and (c,f) after white light illumination.*

**Electronic structure, transport calculations and discussion.**

The geometric structures of the TPT derivatives in their open and closed forms were first optimized in the gas phase at DFT level, with the B3LYP functional[41] and a 6-31 G (d,p)[42] basis set with the Gaussian09 software.[43] The relaxed molecules are then tilted to fit the measured SAM thickness when chemisorbed on the Au(111) and Co(111) cobalt surfaces through the sulfur anchoring atom. The unit cell of the Au (Co) surface is modeled by a slab of five layers with 4 × 5 gold (5 × 6 cobalt) atoms in each layer and lattice parameters: a= 11.53 Å (12.56 Å), b= 14.42 Å (15.08 Å) and α= 120°. With one molecule per unit cell, this corresponds to a theoretical area per molecule of 144.04 Å$^2$ (164.10 Å$^2$). A vacuum region of 30 Å is introduced above the surface and 10 Å below it. The geometry of the interface was then optimized by relaxing the molecules forming the SAMs and the top two metal layers until forces are below 0.025 eV/Å. The convergence criterion associated with the Self-Consistent Field (SCF) loop is that the energy difference must be smaller than 2x10$^{-5}$ Ha. For this relaxation, we use the Perdew-Burke-Ernzerhof (PBE) functional within the generalized gradient approximation (GGA) and the spin generalized gradient approximation (SGGA)[44] for Au-TPT and Co-TPT SAMs, respectively as implemented in the QuantumATK software.[45, 46] The valence electrons are described within the LCAO approximation, with a single zeta plus polarization basis set (SZP) for metal atoms and a double zeta plus polarization basis set (DZP) for the other atoms. The core electrons are described by the norm-conserving Troullier-Martins pseudopotentials.[47] A density mesh cutoff of 100 Ha and a (5×5×1) k sampling were used in all relaxations. Once the SAM geometries are optimized, a second gold electrode is added on the top side of the molecular layer to build the Au-TPT/Au and Co-TPT/Au junctions (Figs. 8 and 11). Albeit a PtIr tip is used in the experiments, a gold one is used for the calculations as validated in our previous work.[10] A van der Waals contact is assumed between the molecular layer and the top electrode, with an interatomic distance determined as the sum of the van der Waals radii of the hydrogen and gold atoms (2.86 Å). Noteworthy, a layer of gold (platinum) ghost atoms has been added on the top layer of the gold (cobalt) electrodes at a distance of 1.7 (1.6 Å) away so that the work function of the clean Au(111) (Co(111)) surface of 5.25 (5.07 eV) eV matches the experimental value and previous theoretical studies.[10, 48-51] The transmission spectra of the TPT junctions were calculated by the combination of DFT with the Non-Equilibrium Green's Function method (NEGF),[13] as implemented in QuantumATK Q-2019.12-SP1 (see methods).[52] Finally, the I-V characteristics were calculated on the basis of the Landauer-Büttiker formalism, which links the transmission spectrum to the current in a coherent transport regime (more details in the Supporting Information).[14]

### *Au-TPT/Au molecular junctions.*

The relaxed Au-TPT/Au junctions (Fig. 8) exhibit a tilt angle of ≈50 ± 5°, with a junction thickness between 12.5 Å and 13.63 Å, in good agreement with the ellipsometry data (9-10 ± 2 Å). These geometries are also consistent with the appearance in the XPS spectra of the signature of N atoms interacting with the substrate (the "coordinated-like" peak, *vide supra*) and a larger than expected contribution of S-Au bonds (*vide supra* and in the Supporting Information).

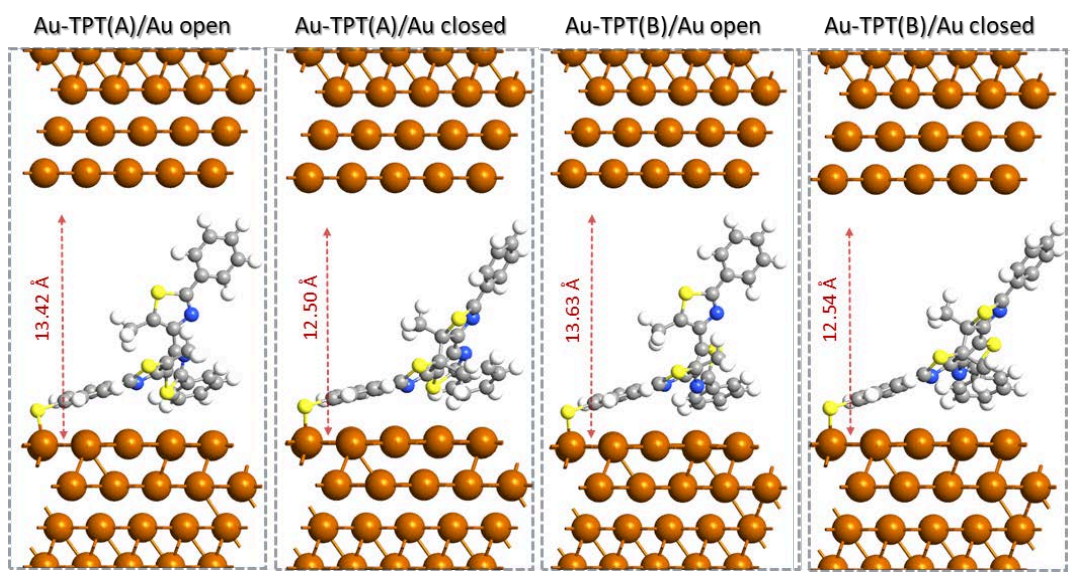

***Figure 8.*** *Optimized Au-TPT/Au junctions. The calculated junction thickness is also marked. The ghost atoms have been removed for better visibility.*

The transmission spectra at zero bias for Au-TPT(A)/Au and Au-TPT(B)/Au junctions are plotted in Fig. 9. For both Au-TPT/Au junctions, the closed isomer exhibits a higher conductance compared to the open isomer as reported in the literature for the diarylethene derivative-based junctions.[11, 12, 53] Interestingly, the theoretical $R^{TC}_{c/o}$ ratio for the transmission coefficient, identified as TC(closed)/TC(open), where TC(closed/open) represents the transmission at the Fermi level through the closed/open form, is larger by a factor of 2.6 for Au-TPT(B)/Au than for Au-TPT(A)/Au junction, in excellent agreement with experiment (factor 2-2.5). Moreover, we notice upon photoswitching from the open to closed form a weak shift of the HOMO level (the closest level with respect to the Fermi level) by 0.22 eV and 0.3 eV for Au-TPT(A)/Au and Au-TPT(B)/Au, respectively. This is in line with the analysis of the experimental datasets based on the SEL model pointing to a weak shift of the energy $\varepsilon_0$ of the level responsible for the transport (Figs. 6, 5 and Table 2), albeit more pronounced compared to the experiments (shift of ≈ 0.1 - 0.18 eV, *vide supra*). Besides the small shift of the HOMO towards the Fermi level in the closed form, the simulated transmission spectra also reveal that the higher conductance of the closed forms also originates from a stronger coupling with the Au electrode, as reflected by the larger width of the resonance peak compared to open isomers. The fitted broadening of the peaks of the transmission spectrum, Γ, of 100 meV for the closed form is larger than the value of 45 meV for the open isomer, see Fig. S14 in the Supporting Information, yielding an increase in the transmission in the HOMO-LUMO gap for the closed form. We note that this broadening Γ is expected to be very sensitive to the top electrode contact configuration (the molecular level broadening Γ is the sum $Γ_1 + Γ_2$). As this top electrode contact can only be assumed at the theoretical level, the simulated broadening must be discussed at the qualitative level. Moreover, the SEL equations rely on important approximations and can only give estimates of the parameters. At this qualitative level, the SEL analytical model agrees with the simulation, as both shows an increased of the electrode coupling for the closed forms (Figs. 6 and 7). It is worth stressing that the simulation of Au-TPT/Au junctions assuming a non tilted configuration gives similar results and orders of magnitude compared to the tilted one (see Figs. S15 and S16 in the Supporting Information), but without the change in the broadening of the transmission peak between the two forms, thus pointing to the key role played by the interface.

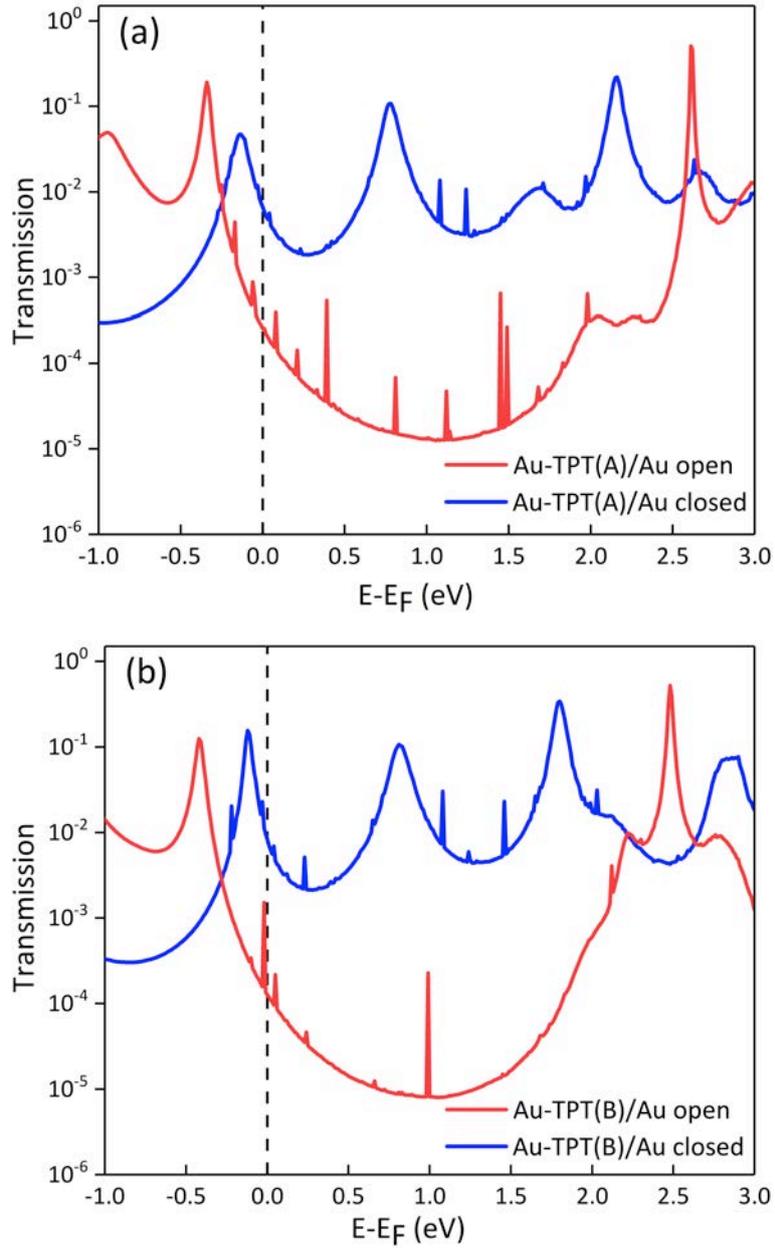

*Figure 9.* Log scale plot of the transmission spectra at zero bias for (a) Au-TPT(A)/Au and (b) Au-TPT(B)/Au junctions in their closed and open forms. The calculated values of $R_{c/o}$ are 26.5 and 69 for TPT(A) and TPT(B), respectively.

For a more realistic comparison between theory and experiments, the I-V curves and the voltage-dependent ratios $R_{c/o}(V)$ were simulated by using the Landauer-Büttiker formula[14] through the integration of the transmission spectrum calculated in a self-consistent way for each bias from 0V to 0.75 V (see Supporting Information). The simulated I-V curves (Fig. 10a) demonstrate that for both Au-TPT(A)/Au and Au-TPT(B)/Au junctions, the closed isomers exhibit higher current values compared to those of the open isomers. The theoretical $R_{c/o}$ ratio here described as I(closed)/I(open), with I(closed) and I(open) the current through the closed and open from, was also calculated (Fig. 10b and Table 3). As predicted from the transmission at zero bias, the Au-TPT(B)/Au junction exhibits higher $R_{c/o}$ compared to Au-TPT(A)/Au by a factor of 2.6 at 0.5 V, in excellent agreement with experiments (i.e. ≈2.3 at 0.5V, Figs 6,7 and Table 1). The magnitudes of the individual $R_{c/o}$ also show the same trends as the experimental values (around 20 and 50 theoretically versus ≈7 and ≈15 experimentally for TPT(A) and TPT(B), respectively). Noteworthy, the higher slope that appears beyond 0.5V in Fig. 10b is explained by the incorporation of the LUMO transmission peak of the closed forms in the transmission window; this cannot occur with the LUMO transmission peak associated with the open form that has higher energy. This feature significantly increases the current value for closed forms and accordingly $R_{c/o}$. This result is consistent with the experimental $R_{c/o}(V)$ values (Fig. S10 especially for the TPT(B)-Au junctions, Supporting Information). It validates the choice to restrict the fits of the SEL model in the voltage window -0.5 to 0.5V to avoid the contribution of a second energy level not included in this model (see section 6, Supporting Information).

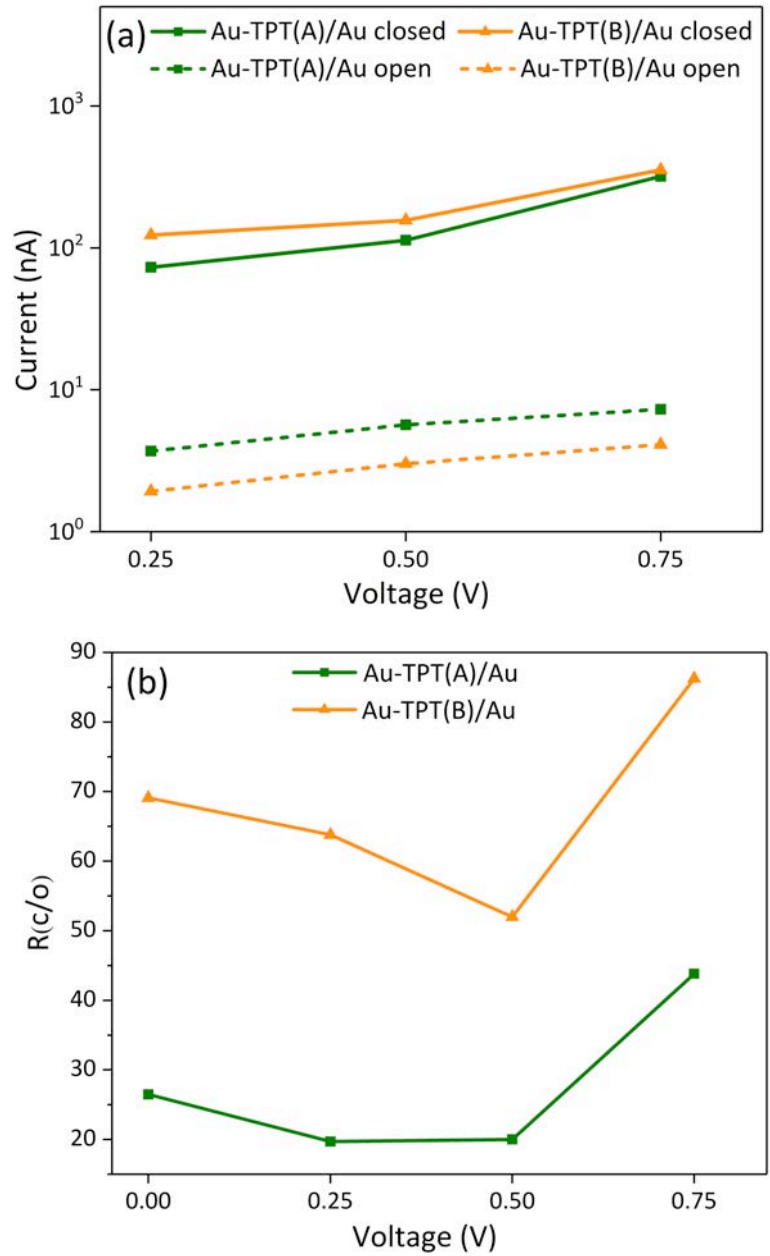

*Figure 10. (a) Current-voltage curves for Au-TPT(A)/Au (green) and Au-TPT(B)/Au (orange) junctions in the open (dashed lines) and closed (solid lines) forms. (b) Corresponding closed/open ratios calculated as a function of the bias.*

| Bias (V) | $R_{c/o}$ | | | |
|---|---|---|---|---|
| | Au-TPT(A)/Au | Au-TPT(B)/Au | Co-TPT(A)/Au | Co-TPT(B)/Au |
| 0.00 | 26.5 | 69.1 | 37 | 288 |
| 0.25 | 19.7 | 63.8 | 39 | 222 |
| 0.50 | 20.0 | 52.0 | 44 | 159 |
| 0.75 | 43.8 | 86.2 | 43 | 113 |

**Table 3.** *Closed/open ratios ($R_{c/o}$) for both Au-TPT/Au and Co-TPT/Au junctions calculated at the Fermi level (0 V) and at several voltages.*

We have plotted the evolution of HOMO level with respect to the average Fermi level of the electrodes as a function of the applied voltage for the Au-TPT/Au junctions (see Figure S17 in the Supporting Information). We found that the offset of the HOMO level tends to increase with the bias and that the open forms exhibit a deeper HOMO level compared to the closed counterparts by an average value of 0.16 eV and 0.22 eV for Au-TPT(A)/Au and Au-TPT(B)/Au junctions, respectively. This agrees well with the fitting done using the SEL, showing a similar voltage-dependent evolution and a higher energy gap between the HOMO resonance and the Fermi level for the open forms compared to the closed ones by 0.10-0.18 eV (see Fig. S11 in the Supporting Information).

***Co-TPT/Au molecular junctions.***

The relaxed Co-TPT/Au junctions displayed in Fig. 11 lead to a geometrical configuration showing a strong interaction of the TPT molecules with the Co electrode, which gives rise to a strong interaction between Co and N, S, C atoms of TPT molecules, consistent with the experimental XPS data (*vide supra* and Supporting Information). We have calculated the total amount of charge transferred at the metal-SAM interface[54-56] for Au-TPT and Co-TPT systems and have found that the Co-TPT SAMs exhibit a higher interfacial charge transfer compared to Au-TPT (a maximum of 0.8|e| for Co-TPT versus 0.37|e| for Au-TPT, see Supporting Information, Figs. S18 and S19), which thus reflects stronger interaction between the TPT molecules and the Co surface.

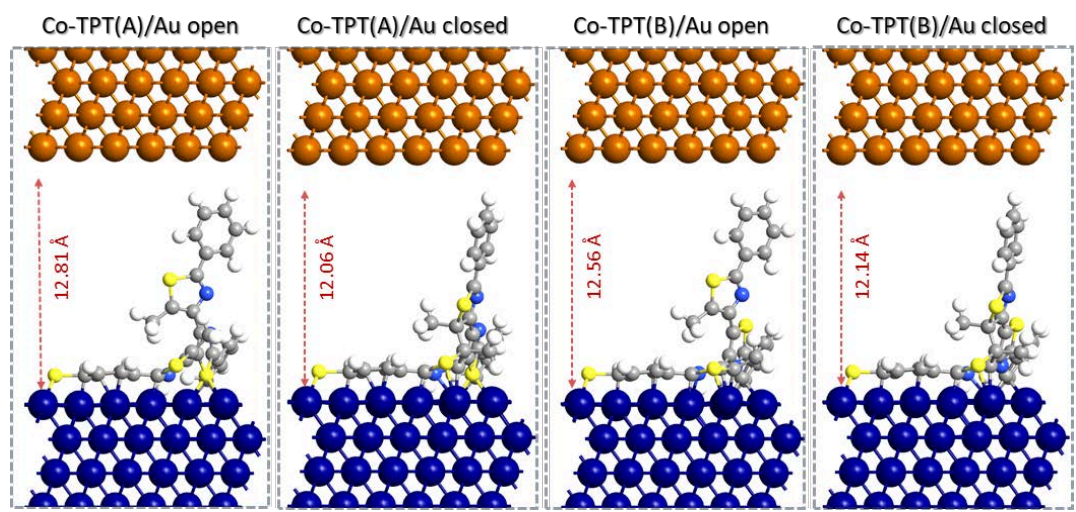

*Figure 11.* Optimized Co-DAE/Au junctions. The calculated junction thickness is also marked. The ghost atoms have been removed for better visibility.

The transmission spectra at zero bias for Co-TPT/Au junctions are illustrated in Fig. 12. The transmission spectra were calculated for spin-up and spin-down electrons (see Fig. S20 in the Supporting Information). However, since we have not performed spin-polarized current measurements, we report in Fig. 12 the total transmission spectra (spin up + spin down). As for Au-TPT/Au junctions, the closed forms exhibit a higher conductance compared to open forms with $R_{c/o}$ ratio of 37 and 288 for Co-TPT(A)/Au and Co-TPT(B)/Au junctions, respectively. We also simulated the I-V curves for the Co-TPT/Au junctions (Fig. 13a) and the bias-dependent $R_{c/o}$ under bias from 0.25 V to 0.75 V in first approximation by integrating the transmission spectra calculated at zero bias (see Supporting Information). This is motivated by the fact that calculations with Co electrodes are much more time consuming due to the need for spin-polarized calculations. We have found by comparing zero-bias vs. finite-bias voltage calculations for the two molecules on Au that the discrepancy associated to the use of the zero transmission spectra for estimating the closed/open ratio magnitude can be sensitive to the bias and the investigated junctions, see section 8 in the supporting information. The trends obtained for Co-TPT/Au junctions at equilibrium might thus evolve with bias, however, these trends, i.e. a higher closed/open ratio for the Co-TPT(B)/Au junction than for the Co-TPT(A)/Au one, are similar, see section 8 in the supporting information. The Co-TPT(B)/Au junction exhibits a higher $R_{c/o}$ than Co-TPT(A)/Au by a factor of 3.6 at 0.5 V (Fig. 13b and Table 3), which is in good agreement with experiment (≈2 at 0.5V).

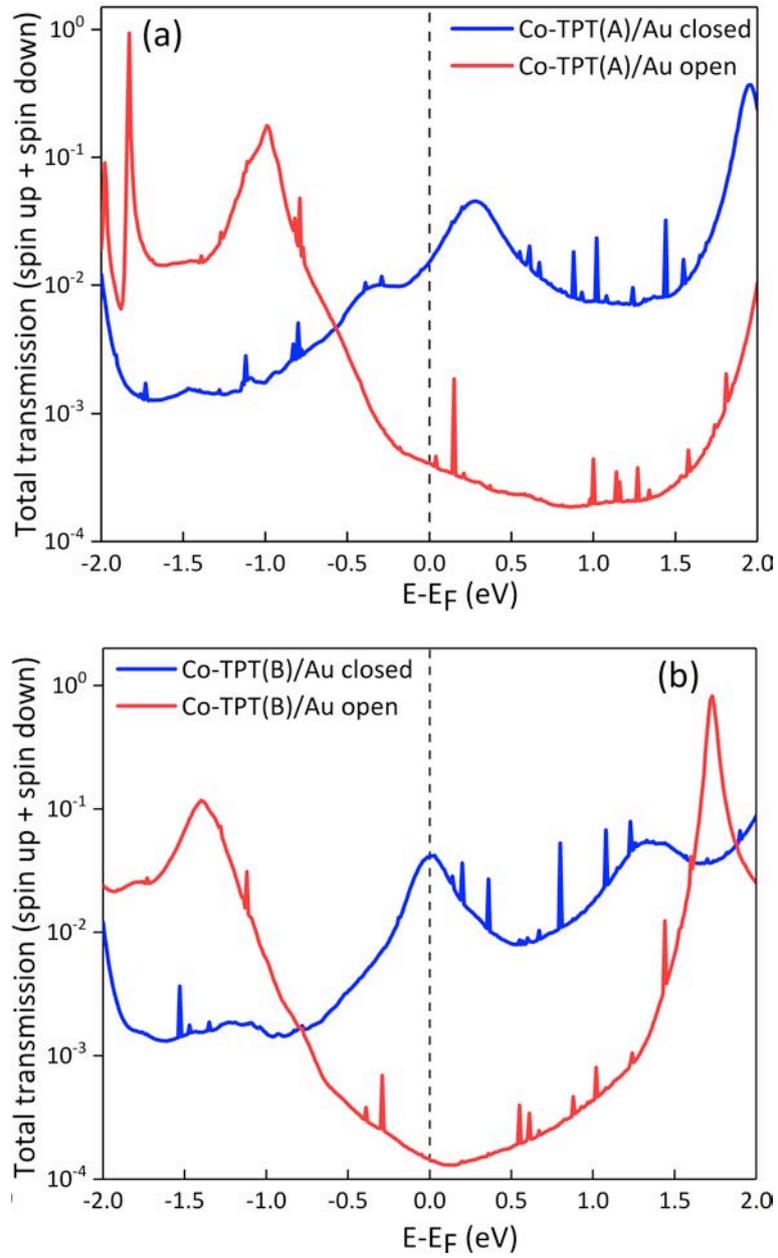

*Figure 12.* Log scale plot of the transmission spectra (spin up + spin down) at zero bias for (a) Co-TPT(A)/Au and (b) Co-TPT(B)/Au junctions in their closed and open forms. The calculated values of $R_{c/o}$ are 37 and 288 for TPT(A) and TPT(B), respectively.

The TPT molecules exhibit higher $R_{c/o}$ on Co compared to the same molecule on Au, particularly TPT(B) for which the calculated $R_{c/o}$ (159 at 0.5V) is as high as the experimental value (average of 285 at |0.5| V, Table 1); the corresponding values are 44 versus 170 for TPT(A), respectively. From a theoretical point of view, the different behavior of the Au and Co junctions is associated to the stronger interfacial interactions with the cobalt electrode. As illustrated by the transmission spectra displayed in Fig. 12, the closest transmission peak with respect to the Fermi level (HOMO for open isomers (red line) and LUMO for closed isomers (blue line)) exhibits a much larger broadening compared to their counterparts on Au (Fig. 9). This is further evidenced by severe shifts of the transmission peaks when going from the open to the closed form. Indeed, the HOMO transmission peak of the open forms is located far from the Fermi level, at -1.00 eV and -1.40 eV for Co-TPT(A)/Au and Co-TPT(B)/Au junctions, respectively. In contrast, the closed forms display LUMO transmission peaks lying close to the Fermi level at 0.28 eV for Co-TPT(A)/Au and in resonance for Co-TPT(B)/Au junctions. The latter peak is thus responsible for the higher $R_{c/o}$ calculated for the Co-TPT(B)/Au junction. This behavior is consistent with a strong coupling of the N atoms of the TPT(B) molecule with the Co surface, as observed by the presence of a "coordinated-like" peak in the XPS measurements (Fig. S5 in the Supporting Information). In particular, the higher [C=N⋯Co]/[N=C] ratio for the TPT(B) SAMs (Table S1 in the Supporting Information) is consistent with the simulations showing that 2 N atoms are in interaction with the Co surface for the TPT(B) molecule, while there is only one N atom for the TPT(A) molecule (Fig. 11, and also see Fig. S20 in the Supporting Information for a closer view). Note that the $R_{c/o}$ decreases with increasing bias for the Co-TPT(B)/Au junction (see Fig. 13b) because the LUMO transmission peak of the closed form is initially already fully incorporated in the transmission window (see Fig. 12b), leading to the saturation of the current while the current of the open form keeps increasing with the increasing bias. This decreasing $R_{c/o}$ with voltage is clearly observed in the experiments for TPT(B) SAMs on Co (Fig. S10 in the Supporting Information). The approximation made by simulating the I-V curves from the transmission spectra calculated at zero bias may explain why TPT(A) does not exhibit in the calculations a $R_{c/o}$ value as high as that found from experiment (155 at 0.5V, Table 1).

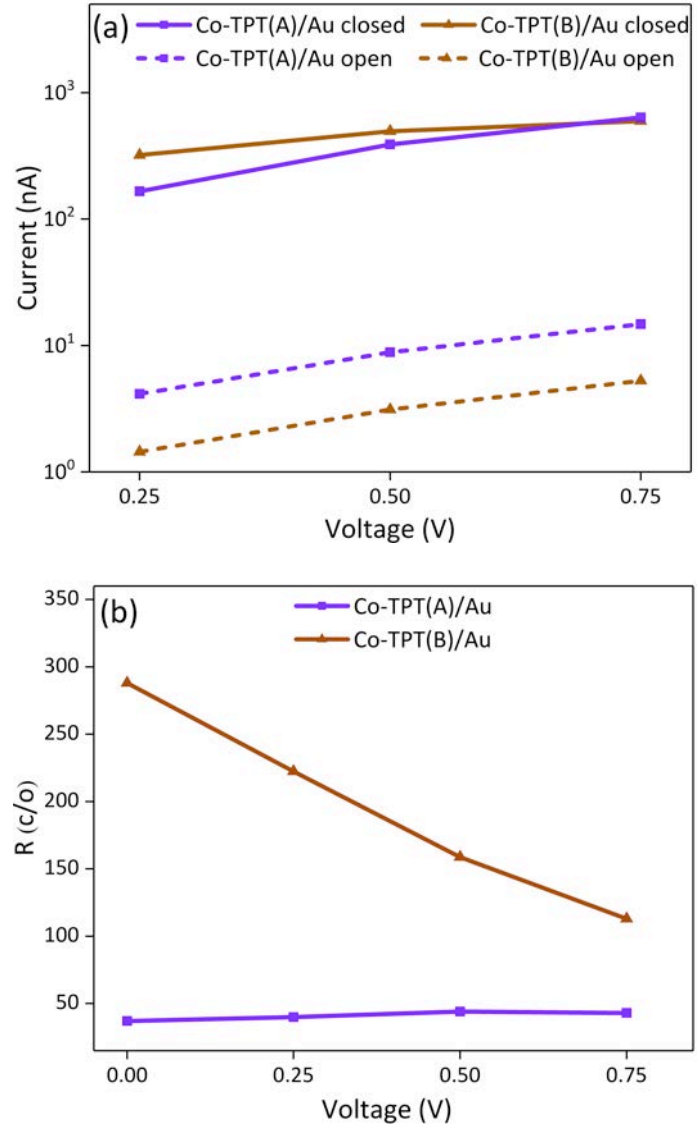

***Figure 13.*** *(a) Current-voltage curves for the Co-TPT(A)/Au (violet) and Co-TPT(B)/Au (brown) junctions in the open (dashed lines) and closed (solid lines) forms. (b) Corresponding closed/open ratios ($R_{c/o}$) calculated as a function of the voltage. The I-V curves and $R_{c/o}$ of the Co-TPT/Au junctions are calculated in first approximation by integrating the transmission at zero bias in different voltage ranges (see Supporting Information).*

It is difficult to plainly explain the stronger coupling of cobalt compared to gold. However, this feature may be originating from the different character and strength of hybridization between the molecular orbitals and the bands of the metal surface. In the case of cobalt, the molecules are strongly chemisorbed through the hybridization between the d-Co states and the p-orbitals of the molecules, whereas a weaker hybridization between the s-Au states and the p-orbitals of the molecules is taking place in the case of a gold surface. Interestingly, several studies have also noticed a strong coupling between organic molecules and cobalt,[57-59] an interesting trend that could be confirmed by investigating other Co-molecule systems.

To further shed light on the conductance of the actual Co-TPT/Au junctions, we have also simulated the transmission spectra in the non tilted configuration (see Fig. S22 in the Supporting Information). The results confirm the previous findings: the Co-TPT(B)/Au junction exhibits $R_{c/o}$ twice as large compared to the Co-TPT(A)/Au junction and the $R_{c/o}$ of TPT is higher on the Co electrode compared to the same molecule on the Au electrode. However, the calculated $R_{c/o}$ (15.7 for Co-TPT(A)/Au and 33.5 for Co-TPT(B)/Au) are not as high as the $R_{c/o}$ values obtained from experiments and the simulated transmission spectra of the lying-down configurations. Moreover, for both Co-TPT(A)/Au and Co-TPT(B)/Au junctions, the simulated transmission spectra of the non tilted configurations (Fig. S23 in the Supporting Information) show a weak energy shift of the HOMO transmission peak upon isomerization (0.20 eV) accompanied by a smaller broadening as a result of the weaker coupling with the Co electrode. Thus this "non tilted" configuration can be discarded to explain the main experimental results. However, this "non tilted" situation has been encountered in another batch of TPT(B) on Co for which the XPS data showed a very weak "coordinated-like" N 1s peak (Fig. S5, batch#2), i.e. weak molecule/electrode coupling, and the CAFM measurements (Fig. S24) yielded a weak value of $R_{c/o}$ of about 15-25. Thus, this "counter experiment" comforts the conclusion that the high values of the $R_{c/o}$ ratios are related to the strong tilt of the molecules allowing the N atoms to interact with the electrode, and that the TPT(B) molecule favors this interaction with the anchoring group (thiol) lying the same side as two N atoms of the thiazole units (instead of only one N atom for the TPT(A) molecule).

## Conclusion.

Molecular junctions of two slightly different terphenylthiazoles on Co electrodes showed high photo-switching conduction ratios between the closed and open forms (150 to 380), about 20 times larger than for the same molecular junctions on Au. The highest ratios are obtained when the thiol group is on the same side as the 2 N atoms of the thiazole units, a situation which favors the coupling of these N atoms with the metal electrode. The experimental results are consolidated and rationalized by first principle calculations. These high ratios are due to a transition from a HOMO mediated off-resonance electron transport (open form) to a LUMO mediated (quasi-) resonant electron transport for the closed form. This behavior (not observed on Au electrodes) is related to the strong coupling and a favorable level alignment for the closed form of the molecules on the Co electrode, highlighting both the role of the chemical nature of the electrode and the effect of interfacial geometry on tuning the transport behavior in molecular junctions.

## Methods

### *Synthesis and sample fabrication.*

*Molecule synthesis.* The two new photochromic terphenylthiazoles (TPT) A and B were prepared according to the synthetic routes described in details in the Supporting Information (section 1). These air-sensitive syntheses were performed under argon using standard Schlenk techniques.

*Bottom metal electrode fabrication.* Template stripped gold ($^{TS}$Au) substrates were prepared according to the method reported by the Whiteside group.[60, 61] In brief, a 300–500 nm thick Au film is evaporated on a very flat silicon wafer covered by its native $SiO_2$ and then transferred to a glued clean glass piece which is mechanically stripped with the Au film attached on the glass piece, letting exposed a very flat (RMS roughness of 0.4 nm, the same as the starting $SiO_2$ surface used as the template). We prepared cobalt substrates by evaporating about 40 nm of cobalt on highly-doped n-Si(100) substrates with a vacuum evaporation system installed inside the glovebox. See more details section 2 in the Supporting Information.

*Self-assembled monolayers*. The SAMs on $^{TS}$Au and Co were fabricated by dipping the freshly prepared metal substrate in a solution of TPT molecules in anhydrous EtOH/THF (80/20) at 0.2mM for 1 day in the dark (section 2 in the Supporting Information).

*Transfer under controlled atmosphere (SAMs on Co only).* The transfer of the samples from the glovebox to the CAFM and the X-ray Photoelectron Spectroscopy (XPS) instrument under UHV was carried out in a homemade hermetic transport container under overpressure of $N_2$.

**Spectroscopic ellipsometry.**

The thickness of the SAMs was measured by spectroscopic ellipsometry (UVISEL ellipsometer (HORIBA), section 4 in the Supporting Information). To avoid the oxidation of the cobalt, the samples were placed in a sealed cell (HORIBA) filled with the $N_2$ atmosphere of the glovebox. TPT SAMs on $^{TS}$Au were measured in ambient conditions.

**XPS measurements.**

XPS experiments were performed to analyze the chemical composition of the SAMs and to check the residual oxidation state of the cobalt surface. We used a Physical Electronics 5600 spectrometer fitted in an UHV chamber with a residual pressure of $3 \times 10^{-10}$ mbar. The measurements were done using standard procedures (section 5 in the Supporting Information).

**CAFM in ambient conditions.** We measured the electron transport properties at the nanoscale by CAFM (ICON, Bruker) at room temperature using a tip probe in platinum/iridium (with loading force of ca. 15 nN). We used a "blind" mode to measure the current-voltage (I-V) curves and the current histograms: a square grid of 10×10 was defined with a pitch of 50 to 100 nm. At each point, the I-V curve is acquired leading to the measurements of 100 traces per grid. This process was repeated several times at different places (randomly chosen) on the sample, and up to several thousand of I-V traces were used to construct the current-voltage histograms (section 6 in the Supporting Information).

**UHV CAFM measurements.**

The CAFM experiments on SAMs under UHV (pressure $10^{-11} - 10^{-9}$ mbar) were performed at room temperature using a VT-SPM microscope (Scienta Omicron). CAFM imaging and local current-voltage (I-V) spectroscopy were carried out using platinum-iridium coated probes. Typically, up to few hundreds of I-V traces were recorded at a loading force of ca. 15 nN (section 6 in the Supporting Information).

**Irradiation setup of the photochromic TPT SAMs.**

A power LED (M365F1 from Thorlabs) was used for the UV light irradiation at 365 nm for the CAFM measurements in air. A chromatographic UV lamp (365 nm) was used for the measurements with the UHV CAFM. For the visible light irradiation, we used a white light halogen lamp (Leica CLS150X). Since the light source to sample distance is not the same for the CAFM in air and in UHV, the irradiation time was adjusted to subject the samples to about the same photon density (section 7 in the Supporting Information).

**Calculations of the electronic structures.**

For the DFT/NEGF calculations with QuantumATK, the exchange-correlation potential is described with the GGA.PBE (SGGA.PBE) functional[44, 46] for Au-TPT/Au (Co-TPT/Au) junctions. The Brillouin zone was sampled with a (5×5×100) k-sampling, a mesh cutoff of 80 Ha and a temperature of 300 K. We expand the valence electrons in a single zeta plus polarization basis set (SZP) for metal atoms and a double zeta polarization basis set (DZP) for the other atoms, except for the platinum ghost atoms that are described by a medium basis set and a PseudoDojo potential. The core electrons are frozen and described by the norm-conserving Troullier-Martins pseudopotentials.[47] These parameters have been carefully tested to ensure the convergence of the computed transmission spectra.

## Associated content

The Supporting Information is available free of charge at xxxxxx: detailed synthesis and RMN characterization of the two new molecules; electrodes and self-assembled monolayers fabrication; UV-vis spectroscopy showing the photo-switching behavior); ellipsometry (SAM thickness); XPS of the monolayers, focussing on the signature of molecule/co coupling; detailed protocols for the C-AFM experiments, data analysis and fits with analytical models; UV and visible light illumination conditions; details of the theoretical calculations (DFT/NEGF) and additional results.

*Author Contributions*

V.P. and D.G. fabricated the SAMs, performed and analyzed the physico-chemical characterizations. V.P. performed all the CAFM measurements (in air and UHV). V.P. and D.V. analyzed the data. A.L., P.Y. and T.M. synthesized the molecules and performed their chemical characterizations. I.A. performed the theoretical calculations under the supervision of C.V.D. and J.C. The project was conceived by D.V., T.M., P.Y., J.C. and supervised by S.L. and D.V. The manuscript was written by D.V. with the contributions of all the authors. All authors have given approval of the final version of the manuscript.

# These authors (V.P. and I.A.) contributed equally to this work.

*Note*

The authors declare no competing financial interest.

## Acknowledgements.

This work has been financially supported by the French National Research Agency (ANR), project SPINFUN ANR-17-CE24-0004. We acknowledge D. Deresmes for his valuable help with the UHV CAFM instrument, Xavier Wallart for the XPS measurements, Y. Deblock for ellipsometry. The IEMN clean-room fabrication and SPM characterization facilities are partly supported by renatech. The work of I.A. is supported by the Belgian National Fund for Scientific Research (F.R.S.-FNRS) thanks to the project SPINFUN (Convention T.0054.20). We also acknowledge the Consortium des Équipements de Calcul Intensif (CÉCI) funded by the Belgian National Fund for Scientific Research (F.R.S.-FNRS) for providing the computational resources. J.C. is an FNRS research director.

# Terphenylthiazole-based self-assembled monolayers on cobalt with high conductance photo-switching ratio for spintronics

*Vladimir Prudkovskiy,[a] Imane Arbouch,[b] Anne Léaustic,[c] Pei Yu,[c,]\**
*Colin Van Dyck,[b] David Guérin,[a] Stéphane Lenfant,[a]*
*Talal Mallah,[c] Jérôme Cornil,[b,]\* and Dominique Vuillaume.[a,]\**

a) Institute for Electronics Microelectronics and Nanotechnology (IEMN), CNRS, Av. Poincaré, Villeneuve d'Ascq, France.

b) Laboratory for Chemistry of Novel Materials, University of Mons, Place du parc 20, 7000 Mons, Belgium.

c) Institut de Chimie Moléculaire et des Matériaux d'Orsay (ICMMO), CNRS, Université Paris-Saclay, 91405 Orsay Cedex, France

pei.yu@universite-paris-saclay.fr
jerome.cornil@umons.ac.be
dominique.vuillaume@iemn.fr

**SUPPORTING INFORMATION**

## Section 1. Synthesis and NMR characterization.

Air-sensitive syntheses were performed under argon using standard Schlenk techniques. Chemicals and solvents were used as received unless otherwise stated. Anhydrous solvents, when necessary, were dried using standard methods. Thin layer chromatography (TLC) was performed on silica gel 60 $F_{254}$ while column chromatography was carried out on silica gel (0.063-0.2 mm). The two new photochromic terphenylthiazoles A and B were prepared according to the synthetic routes shown in Scheme 1.

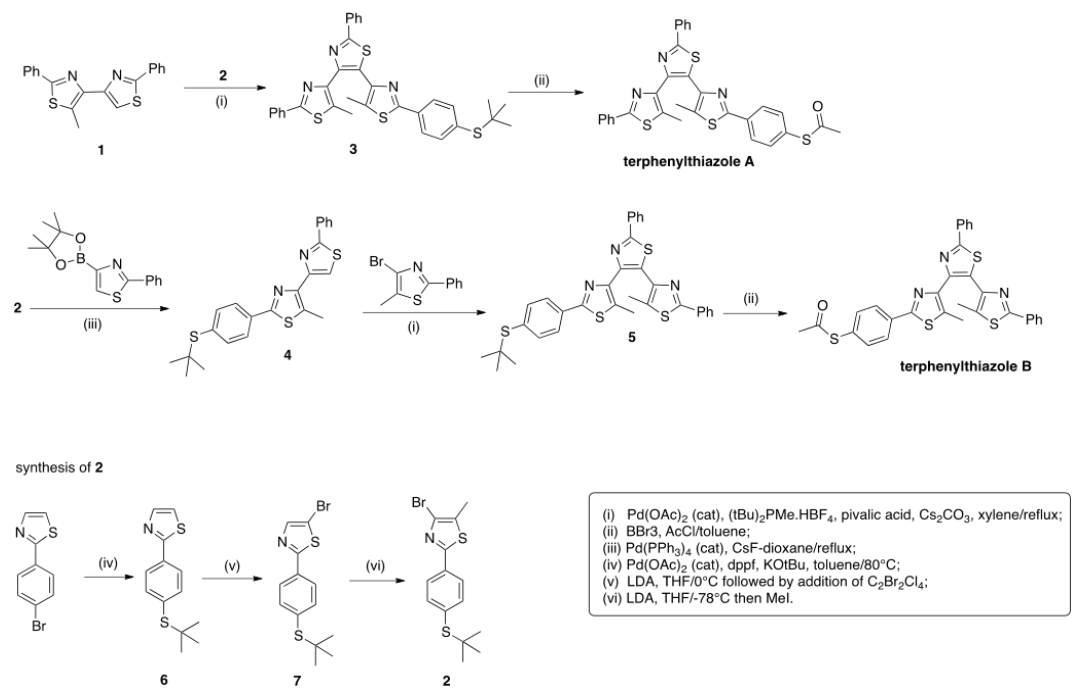

**Scheme S1**

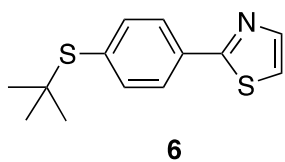

2-(4-Br-phenyl)-thiazole[1] (2.40 g, 10 mmol), KOtBu (1.57 g, 14 mmol), Pd(OAc)2 (23 mg, 0.10 mmol), dppf (60 mg, 0.11 mmol) in a Schlenk were purged before dry toluene (20 ml) and then 2-Methyl-2-propanethiol (1,4 ml, 12.4 mmol) were introduced. The mixture was heated at 80°C under Argon and monitored by TLC until all 2-(4-Br-phenyl)-thiazole was consumed (ca 3 hours). Once cooled to RT water (20 ml) and EtOAc (30 ml) were added to the mixture and the aqueous phase was extracted with EtOAc (2x20 ml). Combined organic phase was washed with water (50 ml) and dried over $Na_2SO_4$. Vacuum evaporation of the solvents gave a light brown crystalline solid (2.375 g, 95%), which is pure enough for the next step or could be further purified by column chromatography (silica gel, dichloromethane) to yield a colorless crystalline solid (2.255 g, yield 90%).

¹HNMR (CDCl₃): 7.95 (d, J = 8.3 Hz, 2H), 7.90 (d, J = 3.5 Hz, 1H), 7.66 (d, J = 8.3 Hz, 2H), 7.37 (d, J = 3.5 Hz, 1H), 1.32 (s, 9H). HRMS (ESI): calculated for $C_{13}H_{15}NS_2$ [M+H]⁺: 250.0719, found: 250.0713.

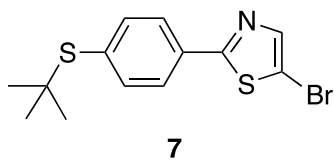

Thiazole 6 (673 mg, 2.7 mmol) was purged before dry THF (20 ml) was introduced and the solution cooled down to 0°C. LDA (2M, 2.0 ml) was added and the resulting deep violet solution was kept at the same temperature for ca 20 min. before 1,2-dibromotetrachloroethane (980 mg, 3 mmol) was added once as solid. The solution turned immediately brown and the solution was warmed to RT and left overnight. Aqueous NH₄Cl solution (1M, 20 ml) and then diethyl ether (20 ml) were added and stirred a few minutes before the organic phase was collected. The aqueous phase was extracted with diethyl ether (20 ml) and combined organic phase was washed with water (40 ml) and dried over Na₂SO₄. After evaporation of the solvents, the solid residue was submitted to column chromatography (silica gel, dichloromethane) to yield thiazole 7 as a light yellow, crystalline solid (755 mg, yield 85%).

¹HNMR (CDCl₃): 7.82 (d, J = 8 Hz, 2H), 7.75 (s, 1H), 7.59 (d, J = 8 Hz, 2H), 1.32 (s, 9H). HRMS (ESI): calculated for $C_{13}H_{15}BrNS_2$ [M+H]⁺: 327.9824, found: 327.9820.

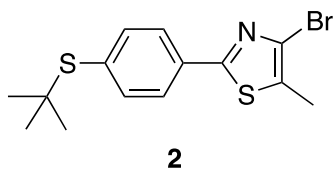

Thiazole 2 is readily accessed through Halogen-Dance reaction.[2, 3] Thiazole 7 (1.315 g, 4 mmol) was purged and dry THF (25 ml) was introduced under argon. The resulting solution was cooled to -78°C before LDA (2M, 3 ml) was added dropwise to give a deep violet solution. After 20 min at the same temperature, iodomethane (0.38 ml, 6 mmol) was added and the solution was slowly warmed to RT and left overnight. To the resulting solution, aqueous $NH_4Cl$ solution (1M, 20 ml) and then diethyl ether (30 ml) were added and stirred a few min. before the organic phase was collected. The aqueous phase was extracted with diethyl ether (20 ml) and combined organic phase was washed with water (40 ml) and dried over $Na_2SO_4$. After evaporation of solvents, the residue was purified by column chromatography (silica gel, dichloromethane) to give thiazole 2 as an off-white crystalline solid (1.250 g, yield 91%).

$^1$HNMR ($CDCl_3$): 7.83 (d, J = 8.2 Hz, 2H), 7.57 (d, 2H), 2.45 (s, 3H), 1.31 (s, 9H).
HRMS (ESI): calculated for $C_{14}H_{17}BrNS_2$ $[M+H]^+$: 341.9980, found: 341.9969.

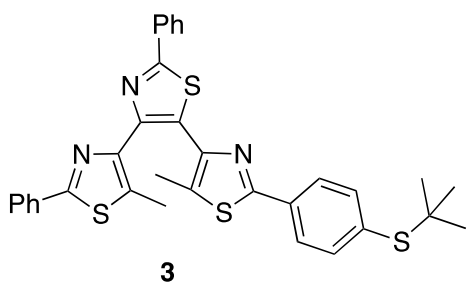

**3**

The title compound was prepared using a reported method via a palladium-catalyzed cross-coupling reaction through a direct C-H activation of thiazole.[4] Bisthiazole 1 [5] (290 mg, 0.87 mmol), thiazole 2 (304 mg, 0.88 mmol), pivalic acid (34 mg, 0.33 mmol), $P(tBu)_2Me \cdot HBF_4$, 30 mg, 0.12 mmol), $Pd(OAc)_2$ (23 mg, 0.1 mmol) and $Cs_2CO_3$ (567 mg, 1.74 mmol) were purged before xylene (5 ml) was introduced under argon. The mixture was refluxed overnight under argon, and dichloromethane (20 ml) then water (20 ml) were added into the mixture at RT. The aqueous phase was extracted with dichloromethane (4x20 ml) and combined organic phase was washed with water (40 ml) and dried over $Na_2SO_4$. After evaporation of the solvents, the residue was purified by column chromatography (silica gel, dichloromethane). The photochromic fraction was evaporated under reduced pressure and the residue was taken in methanol (10 ml) and triturated. The resulting solid was filtered and washed with methanol and dried under vacuum to give terphenylthiazole 3 a light yellow solid (448 mg, yield 86%).

¹HNMR (CDCl$_3$): 8.05-8.08 (m, 2H), 7.89 (d, J = 8.6 Hz, 2H), 7.77-7.80 (m, 2H), 7.58 (d, J = 8.6 Hz, 2H), 7.46-7.48 (m, 3H), 7.33-7.35 (m, 3H), 2.55 (s, 3H), 2.13 (s, 3H), 1.32 (s, 9H). HRMS (ESI): calculated for $C_{33}H_{30}N_3S_4$ [M+H]$^+$: 596.1317, found: 596.1300.

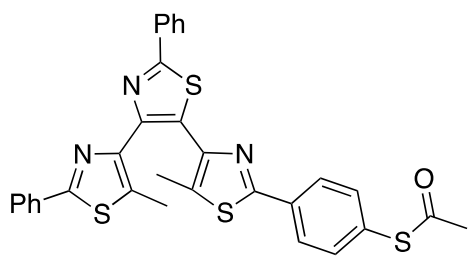

**terphenylthiazole A**

The deprotection of terbutyl thiol and its protection by acetyl group was carried out using BBr$_3$ and AcCl according to known literature method.[6] Terphenylthiazole 3 (300 mg, 0.5 mmol) was purged before addition of dry toluene (5 ml) followed by acetyl chloride (300 μl, 4.2 mmol). BBr$_3$ solution (1M in dichloromethane, 2.6 ml) was added under Argon at 0°C and the mixture stirred at that temperature then at RT overnight. Water (10 ml) was slowly added to destroy the excess of BBr$_3$ and mixture was then extracted with dichloromethane (2x20 ml). Combined organic phase was washed with brine (30 ml) and dried over Na$_2$SO$_4$. After evaporation of the solvent, the residue was purified by column chromatography (silica gel, DCM to 2-3% Et$_2$O) and the fraction containing terphenylthiazole 3 was evaporated under reduced pressure to give a gum-like solid, which was taken up with MeOH (10 ml) and stirred at RT to give the title compound as an off-white solid (188 mg, yield 65%).

1HNMR (CDCl$_3$): 8.05-8.08 (m, 2H), 7.97 (d, J = 8.6 Hz, 2H), 7.77-7.80 (m, 2H), 7.46-7.49 (m, 5H), 7.34-7.36 (m, 3H), 2.55 (s, 3H), 2.45 (s, 3H), 2.12 (s, 3H).

13CNMR (CDCl$_3$): 193.63, 167.30, 164.01, 163.01, 147.70, 146.19, 144.04, 134.76, 134.31, 133.57, 133.55, 133.09, 130.24, 129.69, 129.61, 128.93, 128.78, 126.94, 126.65, 126.29, 30.38, 12.75, 12.41.

HRMS (ESI): calculated for C$_{31}$H$_{24}$N$_3$OS$_4$ [M+H]$^+$: 582.0797, found: 582.0777.

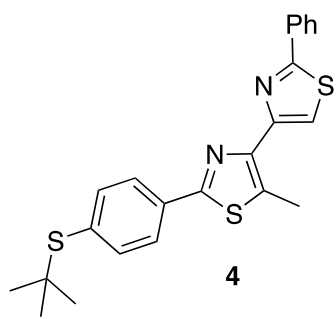

Thiazole 4 was prepared by Suzuki cross-coupling between thiazole 2 and 2-phenyl-4-(4,4,5,5-tetramethyl[1,3,2]dioxaborolan-2-yl)-thiazole[7] (2-Ph-4-Bpin-thiazole). Thiazole 2 (512 mg, 1.5 mmol), 2-Ph-4-Bpin-thiazole (525 mg, 1.83 mmol), Pd(PPh$_3$)$_4$ (72 mg, 0.062 mmol) and CsF (570 mg, 3.75 mmol) were purged before anhydrous dioxane (30 ml) was introduced under argon. The mixture was heated and refluxed under Argon for ca 6 hours and cooled to RT. Water (30 ml) and chloroform (30 ml) were added and organic phase collected. The aqueous phase was extracted with chloroform (25 ml), and combined organic phase was washed with water (50 ml) and dried over Na$_2$SO$_4$. Evaporation of the solvents led to a brown oil, to which MeOH (10 ml) was added and stirred at RT until a crystalline solid was formed. After filtration and washing with MeOH bisthiazole 4 was obtained as a slightly blueish solid (due to the presence of a very few closed form) (550 mg, 87% yield), which was pure enough for the next step.

$^1$HNMR (CDCl$_3$): 8.11 (s, 1H), 8.02-8.05 (m, 2H), 7.99 (d, J = 8.6 Hz, 2H), 7.63 (d, J = 8.1 Hz, 2H), 7.46-7.48 (m, 3H), 3.00 (s, 3H), 1.33 (s, 9H). HRMS (ESI): calculated for C$_{23}$H$_{23}$N$_2$S$_3$ [M+H]$^+$: 423.1018, found: 423.0997.

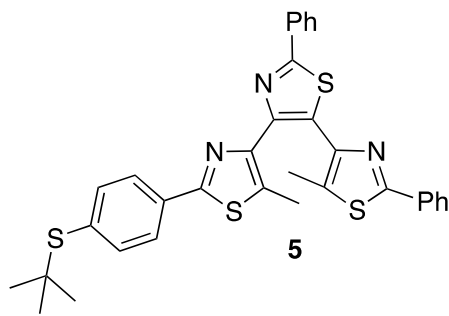

Thiazole 5 was synthesized in a similar way as terphenylthiazole 3. 2-Phenyl-4-Bpin-5-Methyl-thiazole[8] (280 mg, 1.1 mmol), thiazole 4 (405 mg, 1.1 mmol), pivalic acid (43 mg, 0.38 mmol), P(tBu)$_2$Me.HBF$_4$, (38 mg, 0.15 mmol), Pd(OAc)$_2$ (30 mg, 0.13 mmol) and Cs$_2$CO$_3$ (720 mg, 2.2 mmol) were purged before xylene (7 ml) was introduced under argon. The mixture was refluxed overnight under argon, and dichloromethane (30 ml) then water (30 ml) were added into the mixture at RT. The aqueous phase was extracted with dichloromethane (2x25 ml) and combined organic phase was washed with water (40 ml) and dried over Na$_2$SO$_4$. After evaporation of the solvents and the solid residue was stirred with MeOH (15 ml) overnight to give thiazole 5 as an off-white solid, which was used for the next step without further purification (600 mg, 91% yield).

$^1$HNMR (CDCl$_3$): 8.06-8.08 (m, 2H), 7.95-7.97 (m, 2H), 7.73 (d, J = 8 Hz, 2H), 7.42-7.49 (m, 8H), 2.59 (s, 3H), 2.18 (s, 3H), 1.29 (s, 9H). HRMS (ESI): calculated for C$_{33}$H$_{30}$N$_3$S$_4$ [M+H]$^+$: 596.1317, found: 596.1293.

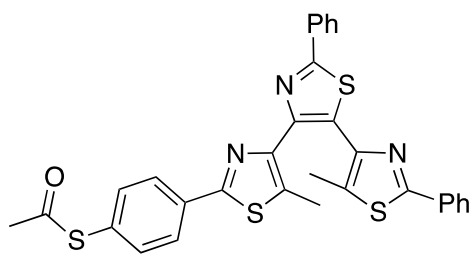

**terphenylthiazole B**

Terphenylthiazole 5 (302 mg, 0.5 mmol) was purged before addition of dry toluene (5 ml) followed by acetyl chloride (300 µl, 4.2 mmol). BBr$_3$ solution (1M in dichloromethane, 2.6 ml) was added under Argon at 0°C and the mixture was then stirred at RT overnight. Water (10 ml) was slowly added to destroy the excess of BBr$_3$ and mixture was then extracted with dichloromethane (2X20 ml). Combined organic phase was washed with brine (30 ml) and dried over Na$_2$SO$_4$. After evaporation of the solvent, the residue was purified by column chromatography (silica gel, DCM to 2-3% Et$_2$O) and the fraction containing terphenylthiazole 3 was evaporated under reduced pressure to give a greenish oil, which was taken up with MeOH (10 ml) and stirred at RT to give the title compound as light cream solid (206 mg, yield 70%).

$^1$HNMR (CDCl$_3$): 8.05-8.08 (m, 2H), 7.92-7.95 (m, 2H), 7.82 (d, J = 8 Hz, 2H), 7.37-7.48 (m, 8H), 2.57 (s, 3H), 2.43 (s, 3H), 2.11 (s, 3H).

$^{13}$CNMR (CDCl$_3$): 193.57, 167.05, 164.13, 162.34, 147.48, 146.45, 143.63, 134.66, 134.53, 133.83, 133.54, 133.41, 132.39, 130.23, 129.98, 129.25, 128.91, 126.89, 126.59, 30.31, 12.80, 12.35.

HRMS (ESI): calculated for C$_{31}$H$_{24}$N$_3$OS$_4$ [M+H]$^+$: 582.0797, found: 582.0770.

## Section 2. Electrodes and SAMs fabrication.

### *General conditions of the process.*

To prevent oxidation of the cobalt substrates, all the preparation of samples (i.e. from metal deposition to grafting of SAMs) was carried out in a nitrogen

MBRAUN glovebox ($H_2O$ and $O_2$ levels below 5 ppm). The glassware was oven dried at 120°C overnight before insertion inside the glovebox to remove residual adsorbed water. The solvents used for the preparation of SAMs (absolute ethanol, tetrahydrofuran) were all purchased anhydrous from Sigma Aldrich. They were stored for 5 days over freshly activated 4 Å molecular sieves (activation for 18h at 150°C under vacuum), then they were degassed with nitrogen for at least 15 min.

### *Bottom metal electrode fabrication.*

Ultraflat template-stripped gold surfaces ($^{TS}$Au), with rms roughness of ~0.4 nm were prepared according to the method developed by the Whitesides group.[9, 10] In brief, a 300–500 nm thick Au film was evaporated on a very flat silicon wafer covered by its native $SiO_2$ (rms roughness of ~0.4 nm), which was previously carefully cleaned by piranha solution (30 min in 7:3 $H_2SO_4/H_2O_2$ (v/v); **Caution**: Piranha solution is a strong oxidizer and reacts exothermically with organics), rinsed with deionized (DI) water, and dried under a stream of nitrogen. Clean 10x10 mm pieces of glass slide (ultrasonicated in acetone for 5 min, ultrasonicated in 2-propanol for 5 min, and UV irradiated in ozone for 10 min) were glued on the evaporated Au film (UV-polymerizable glue, NOA61 from Epotecny), then mechanically peeled off providing the $^{TS}$Au film attached on the glass side (Au film is cut with a razor blade around the glass piece).

Cobalt substrates were prepared by evaporating about 40 nm of cobalt on cleaved (12×10 mm) highly phosphorus-doped n-Si(100) substrates, resistivity < $5.10^{-3}$ Ω.cm (from Siltronix), covered by native oxide, cleaned by 5 min sonication in acetone and isopropanol, then rinsed with isopropanol and dried under $N_2$ flow. The evaporation of 99.99% purity cobalt pellets (Neyco) was realized by Joule effect in a vacuum evaporation system (Edwards Auto306) placed inside the glovebox. The cobalt deposition was realized under a $10^{-6}$ mbar vacuum and at a low rate deposition between 2 and 5 Å/s in order to minimize roughness.

***Self-assembled monolayers of TPT.***

SAMs of the open forms of TPT(A) and TPT(B) on gold and cobalt were prepared from the acetyl-protected thiols by spontaneous assembly at metal surfaces via Au-S or Co-S bonds (see XPS section), without deprotection. Indeed thioacetates are known to undergo spontaneous deprotection at various metal surfaces like gold or silver.[11, 12] The freshly peeled off $^{TS}$Au samples were immediately immersed in millimolar solutions of TPT in anhydrous ethanol/THF (80:20 v/v) for 3 days in the dark. This solvent mixture was compatible with the $^{TS}$Au glue. Then samples were rinsed with degassed anhydrous THF and dried under $N_2$ stream. In a glovebox, the freshly evaporated Co substrates were immediately immersed in millimolar solutions of TPT in anhydrous ethanol/THF (80:20 v/v) for 1 day in the dark. Then samples were rinsed with degassed anhydrous ethanol, dried under $N_2$ stream and stored in the glovebox.

## Section 3. UV-vis spectroscopy.

The reversible isomerization "open to close" of TPT molecules was checked by UV-vis spectroscopy in solution (~µM in $CH_2Cl_2$). UV-Vis absorption spectra were recorded on a Lambda 800 Perkin-Elmer spectrometer. For the UV irradiation of the solutions, we used a 365 nm chromatography lamp (Vilbert Lourmat, light intensity : 2 mW/cm$^2$ at 1 cm distance). For visible irradiation we used a halogen lamp (LEICA model CLS 150X) centered at 600 nm (light intensity : 220 mW/cm$^2$ at 1 cm distance). This experiment performed on µM solutions of TPT(A) and TPT(B) in DCM shows (Fig. S1) that, as expected, irradiation at 365 nm produces a strong absorption peak centered at 600-630nm corresponding to the formation of the closed form with the π-conjugation extended throughout the molecule. The photostationary state is reached after ~10 min of UV irradiation. Then irradiation at 650 nm causes disappearance of the 600-630 nm band and return to the open form. Return to the initial conditions is achieved by a short irradiation in visible light (10-20s). The reversibility, tested for TPT(A), is particularly good after several irradiation cycles (inset Fig. S1).

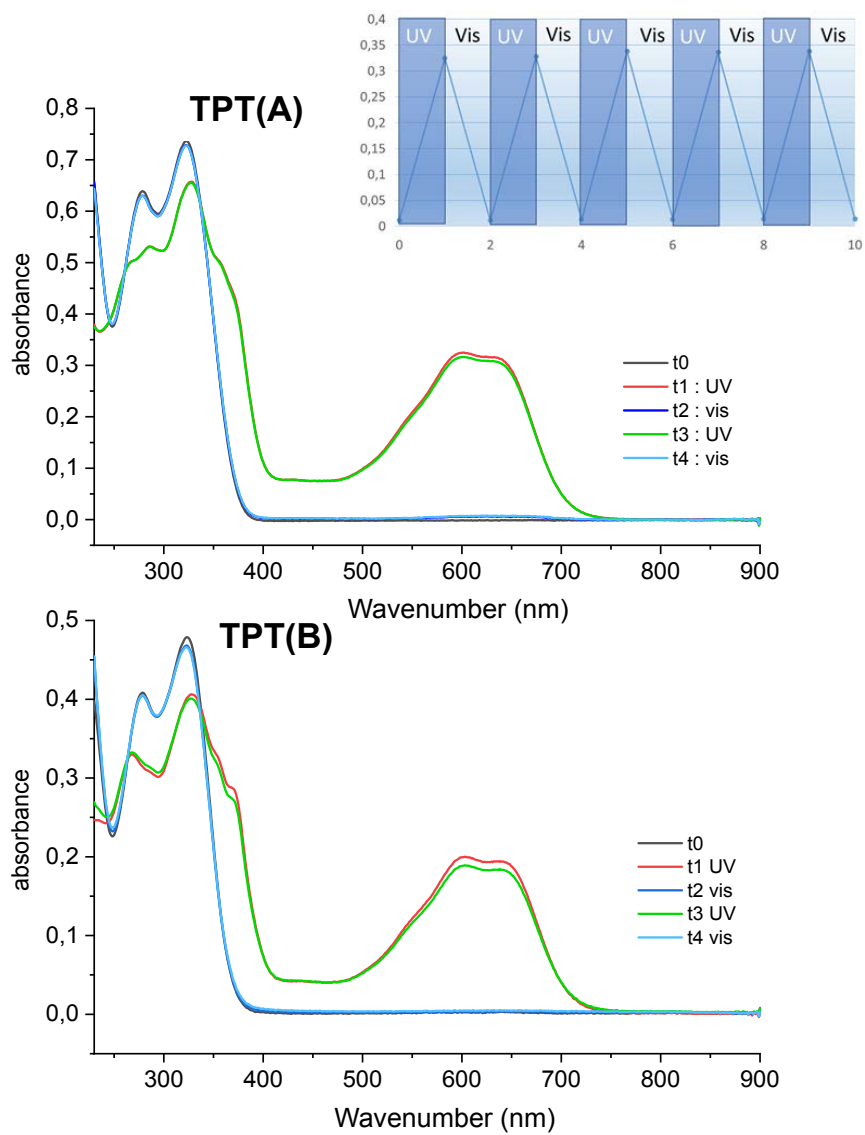

*Figure S1. UV-vis absorbance spectra of TPT(A) (top) and TPT(B) (bottom) in DCM after 4 successive irradiation cycles (2 min for each irradiation step in UV then 2 min in visible light). The pristine state is given at t0. Note that t2 curve is almost not visible here because superimposed on t0 and t4. In the insert at the top right is the absorbance of TPT(A) at 600 nm measured after several irradiation cycles.*

## Section 4. Spectroscopic ellipsometry.

We recorded spectroscopic ellipsometry data (on *ca.* 1 cm² samples) in the visible range using a UVISEL (Horiba Jobin Yvon) spectroscopic ellipsometer equipped with DeltaPsi 2 data analysis software. The system acquired a spectrum ranging from 2 to 4.5 eV (corresponding to 300–750 nm) with intervals of 0.1 eV (or 15 nm). For the measurement in air (SAMs on Au), data were taken at an angle of incidence of 70°, and the compensator was set at 45°. We fit the data by a regression analysis to a film-on-substrate model as described by their thickness and their complex refractive indexes. First, a background for the substrate before monolayer deposition was recorded. We acquired three reference spectra at three different places of the surface spaced of few mm.dSecondly, after the monolayer deposition, we acquired once again three spectra at three different places of the surface and we used a 2-layer model (substrate/SAM) to fit the measured data and to determine the SAM thickness. We employed the previously measured optical properties of the substrate (background), and we fixed the refractive index of the organic monolayer at 1.50.[13] We note that a change from 1.50 to 1.55 would result in less than a 1 Å error for a thickness less than 30 Å. The three spectra measured on the sample were fitted separately using each of the three reference spectra, giving nine values for the SAM thickness. We calculated the mean value from this nine thickness values and the thickness incertitude corresponding to the standard deviation. Overall, we estimated the accuracy of the SAM thickness measurements at ± 2 Å.[14] For SAM on Co using the cell filled with N₂, data were taken at an angle of incidence of 60 ± 1° while the compensator was set at 45°. However, due to the rough Co surface and the use of the cell, the fits with a fixed angle of incidence at 60°C systematically give low values of thicknesses. The fits with this angle as an additional fit parameter give higher values. Consequently, the thicknesses are given with a larger uncertainty (error bar) compared to SAM on Au.

## Section 5. XPS measurements.

High resolution XPS spectra were recorded with a monochromatic Al$_{K\alpha}$ X-ray source (hυ = 1486.6 eV), a detection angle of 45° as referenced to the sample surface, an analyzer entrance slit width of 400 μm and with an analyzer pass energy of 12 eV. In these conditions, the overall resolution as measured from the full-width half-maximum (FWHM) of the Ag 3d5/2 line is 0.55 eV. Background was subtracted by the Shirley method.[15] The peaks were decomposed using Voigt functions and a least squares minimization procedure. Binding energies were referenced to the C 1s BE, set at 284.8 eV.

*TPT(A) and TPT(B) SAMs on $^{TS}$Au.*

The C 1s peak at 284.8 eV (Fig. S3) corresponds to C-C, C-N and C-S bonds. The shoulder observed at 286.2 eV is assigned to the three C=N-S bonds.[16] The S 2p region shows two doublets (S 2p$_{1/2}$ and S 2p$_{3/2}$) associated to the S-C (S 2p$_{1/2}$ at 165.4 eV, S 2p$_{3/2}$ at 164.3 eV) and S-Au (S 2p$_{1/2}$ at 162.9 eV, S 2p$_{3/2}$ at 161.8eV) bonds (Fig. S3). These doublets are separated by 1.2 eV as expected with an amplitude ratio [S 2p$_{1/2}$]/[ S 2p$_{3/2}$] of 1/2. The amplitude ratios [S-Au]/[S-C] are 0.42 for TPT(A) and 0.5 for TPT(B), slightly higher than the 1/3 expected ratio. A small peak at higher energy (≈ 168 eV) is associated to oxidized sulfur (SO$_x$). For both molecules, the N 1s region (Fig. S5) shows two peaks corresponding to "pyridine-like" nitrogen (C=N) and "coordinated-like" nitrogen (C=N...Au) at 398.6 eV and 400 eV, respectively.[17] The piridinic and coordinated-like designations are often used in the literature to describe the components of N 1s signals in N-doped carbons.[18] This N 1s peak splitting is observed when the N atoms interact with a metal surface.[19] The ratio of the peak amplitudes [C=N...Au]/[N=C] is higher for TPT(B) (see Table S1) than for TPT(A) indicating that more N atoms are interacting with the metal electrode for TPT(B) SAM than for the TPT(A) SAM.

**TPT(A) and TPT(B) SAMs on Co.**

The XPS spectra of TPT(A) and TPT(B) on Co show all the C, N and S elements. As for the molecules on $^{TS}$Au, the major peak at 284.8 eV is composed of C-C, C-N and C-S components and a shoulder observed at 286.2 eV is assigned to the three S-C=N carbons (Fig. S4). In the S 2p region (Fig. S4), we observe the contribution of S-C and S-Co bonds (S $2p_{3/2}$ at 164.3 eV, S $2p_{1/2}$ at 165.4 eV for S-C and S $2p_{1/2}$ at 163.7 eV, S $2p_{3/2}$ at 162.6 eV for S-Co). As for the SAMs on Au, the amplitude ratios [S-Co]/[S-C] ~ 0.5-0.6 are higher than the expected 1/3 ratio. Albeit the protocol and precautions used during the grafting and measurements, the O 1s region reveals a residual oxidized Co[20] as in our previous work on azobenzene derivatives on Co (Fig. S3 in Ref. 21). The N 1s region (Fig. S5) shows the two peaks of the C=N bonds (398.6 eV) and the C=N...Co one (400 eV)[17] with ratios of amplitude [C=N...Co]/[N=C] larger for the TPT(B) than for TPT(B) molecules (Table S1). As for the SAMs on Au, this may be due to interaction of N with the surface (large molecule tilt). However, we have also observed (in another batch #2) a case with a low [C=N...Co]/[N=C] ratio which was inferred to a "non tilted" molecule configuration (see discussion section in the main text) for which the N atoms are away from the surface and consequently, only the pyridinic form N=C is detected by XPS.

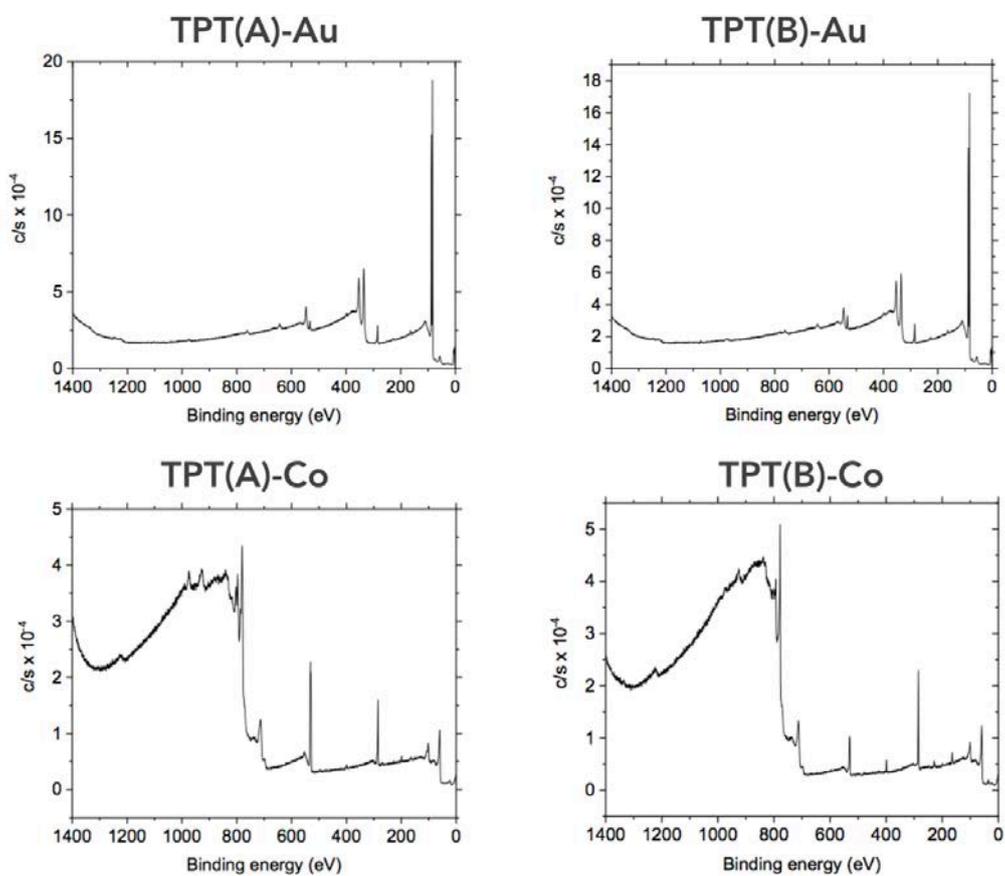

*Figure S2.* XPS survey spectra of the TPT SAMs on Au and Co.

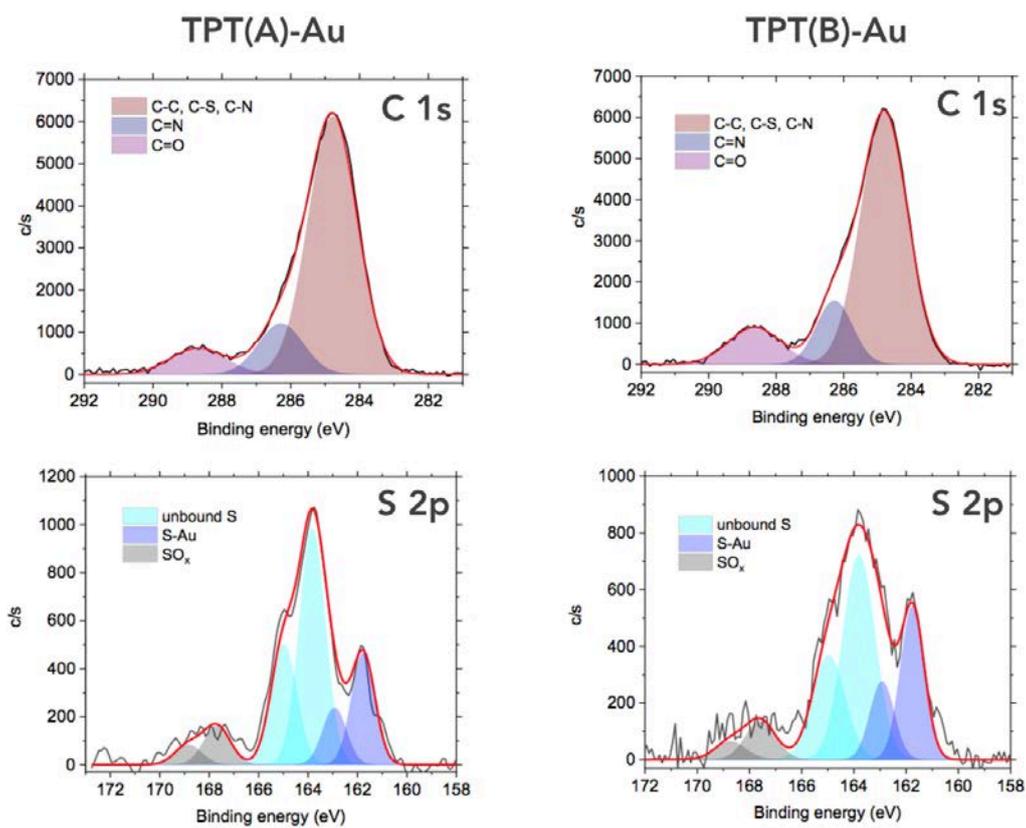

***Figure S3.*** *XPS spectra of the TPT(A)-Au and TPT(B)-Au samples: C 1s and S 2p regions.*

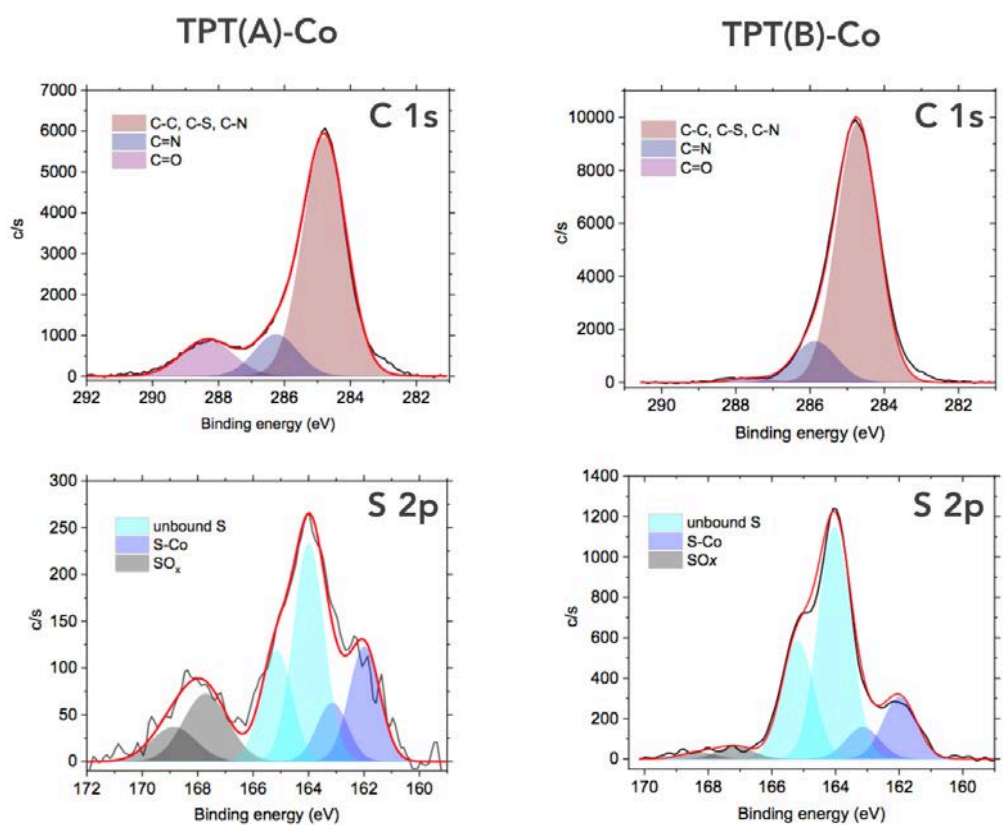

**Figure S4.** XPS spectra of the TPT(A)-Co and TPT(B)-Co samples: C 1s and S 2p regions.

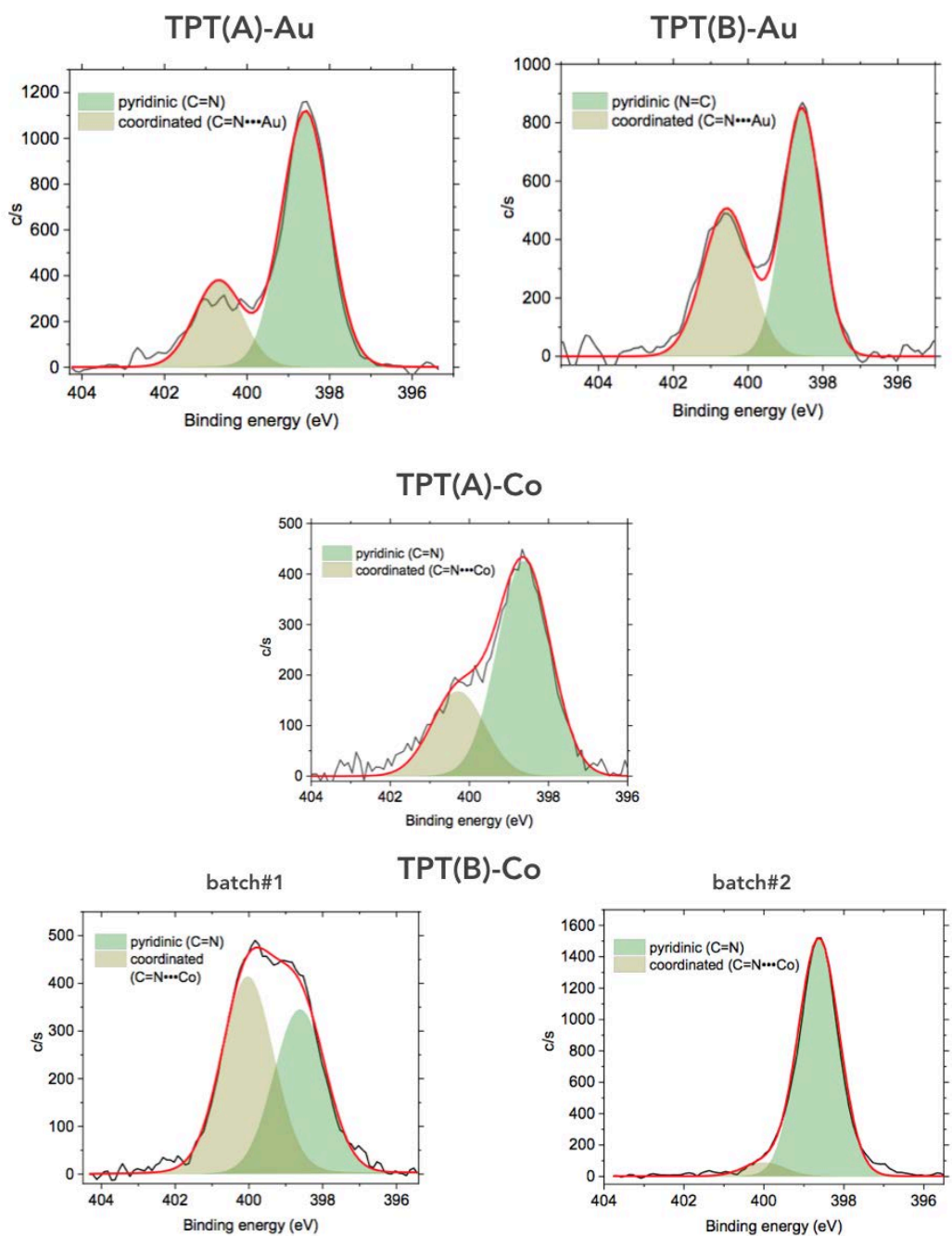

*Figure S5*. XPS spectra of the TPT(A) and TPT(B) SAMs on Au and Co sample: N 1s regions.

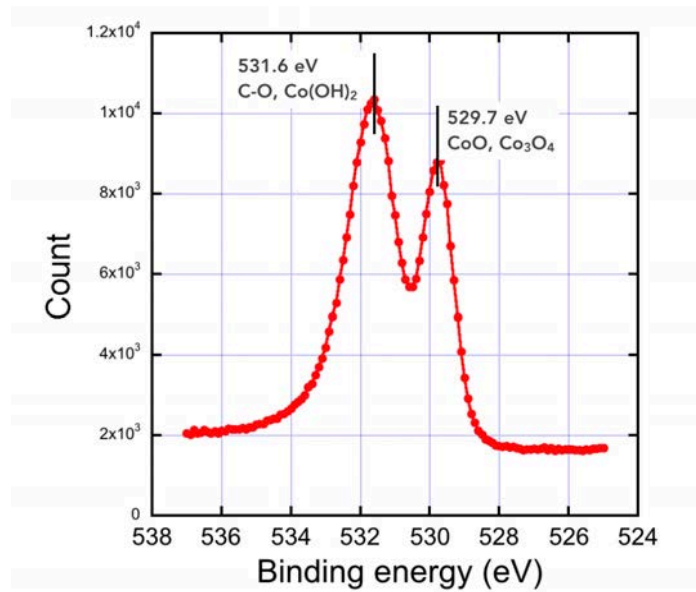

*Figure S6.* XPS spectra of theTPT(B) SAMs on Co sample: O 1s regions. The two peaks are mainly assigned to oxidized Co.[20]

|  | Pyridinic-like Area | Coordinated-like Area | [C=N...M]/[N=C] | BE (eV) |
|---|---|---|---|---|
| **TPT(A)-Au** | 1647 | 555 | 0.34 | 398.6, 400.7 |
| **TPT(B)-Au** | 1085 | 807 | 0.74 | 398.6, 400.6 |
| **TPT(A)-Co** | 719 | 283 | 0.39 | 398.6, 400.3 |
| **TPT(B)-Co #1** | 609 | 757 | 1.24 | 398.6, 400.0 |
| **TPT(B)-Co #2** | 1939 | 111 | 0.06 | 398.6, 400.0 |

*Table S1.* Area of the two N 1s peaks, ratio of the peak amplitude and binding energy of the two peaks (M = Au or Co).

## Section 6. CAFM measurements.

### *CAFM in air.*

Current–voltage characteristics were measured by conductive atomic force microscopy (Icon, Bruker), using PtIr coated tip (SCM-PIC from Bruker, 0.2 N/m spring constant). To form the molecular junction, the conductive tip was located at a stationary contact point on the SAM surface at controlled loading force (∼ 15 nN). The voltage was applied on the substrate. The CAFM tip is located at different places on the sample (typically on an array of stationary contact points spaced of 50-100 nm), at a fixed loading force and the I–V characteristics were acquired directly by varying voltage for each contact point. The I-V characteristics were not averaged between successive measurements and typically between few hundreds and a thousand I-V measurements were acquired on each sample.

### *CAFM in UHV.*

CAFM in UHV ($10^{-11}$ - $10^{-9}$ mbar) were carried out at room temperature with a VT-SPM microscope (Scienta Omicron) and using PtIr coated probes SCM-PIC-V2 (Bruker), tip radius R = 25 nm, spring constant k = 0.1 N/m. In all our measurements, bias (V) was applied on the substrate and the current was recorded with an external DLPCA-200 amplifier (FEMTO). Hundreds to thousands IV traces were acquired using the same protocol as for CAFM measurements in air.

### *Loading force and CAFM tip contact area.*

The load force was set at ∼ 15 nN for all the I-V measurements, a lower value leading to too many contact instabilities during the I-V measurements. Albeit larger than the usual load force (2-5 nN) used for CAFM on SAMs, this value is below the limit of about 60-70 nN at which the SAMs start to suffer from severe degradations. For example, a detailed study (Ref. 22) showed a limited strain-induced deformation of the monolayer (≤ 0.3 nm) at this used load force. The same conclusion was confirmed by our own study comparing mechanical and electrical properties of alkylthiol SAMs on flat Au surfaces and tiny Au nanodots.[23]

Considering: (i) the area per molecule on the surface (as estimated for the thickness measurement and calculated geometry optimization - see theory section), and (ii) the estimated CAFM tip contact surface (see below), we estimate N as follows. As usually reported in literature[22, 24-26] the contact radius, a, between the CAFM tip and the SAM surface, and the SAM elastic deformation, δ, are estimated from a Hertzian model:[27]

$$a^2 = \left(\frac{3RF}{4E^*}\right)^{2/3} \qquad (S1)$$

$$\delta = \left(\frac{9}{16R}\right)^{1/3}\left(\frac{F}{E^*}\right)^{2/3} \qquad (S2)$$

with F the tip load force (15 nN), R the tip radius (25 nm) and E* the reduced effective Young modulus defined as:

$$E^* = \left(\frac{1}{E^*_{SAM}} + \frac{1}{E^*_{tip}}\right)^{-1} = \left(\frac{1-\nu^2_{SAM}}{E_{SAM}} + \frac{1-\nu^2_{tip}}{E_{tip}}\right)^{-1} \qquad (S3)$$

In this equation, $E_{SAM/tip}$ and $\nu_{SAM/tip}$ are the Young modulus and the Poisson ratio of the SAM and C-AFM tip, respectively. For the Pt/Ir (90%/10%) tip, we have $E_{tip}$ = 204 GPa and $\nu_{tip}$ = 0.37 using a rule of mixture with the known material data.[28] These parameters for the DAE SAM are not known and, in general, they are not easily determined in such a monolayer material. Thus, we consider the value of an effective Young modulus of the SAM $E^*_{SAM}$ = 38 GPa as determined for the "model system" alkylthiol SAMs from a combined mechanic and electron transport study.[22] With these parameters, we estimate a = 2 - 2.6 nm (contact area = 13.2 - 21 nm²) and δ = 0.16 - 0.26 nm. With a molecular packing density between 1 to 2 nm²/molecule (as estimated from the tilt angle and theoretical configuration optimization, see theory section), we infer that about 10 molecules are measured in the TPT/PtIr junction, thus we used N=10 in all the I-V fit using Eq. 1 (main text).

*Data analysis.*

Before to construct the current histograms and fit the I-V curves with the one energy-level model, the raw set of IV data is analyzed and some I-V curves were discarded from the analysis:

- At high current, the I-V traces that reached the saturating current during the voltage scan (the compliance level of the trans-impedance amplifier, typically $5 \times 10^{-9}$ A in Figs. S6 and S7, depending on the gain of the amplifier) and/or I-V traces displaying large and abrupt steps during the scan (contact instabilities).

- At low current, the I-V traces that reached the sensitivity limit (almost flat I-V traces) and displayed random staircase behavior (due to the sensitivity limit of both the trans-impedance amplifier and the resolution of the ADC (analog-digital converter), Fig. S7. A typical example of such treatment is shown in Fig. S6. The "measurement yield" for the four samples is summarized in Table S2.

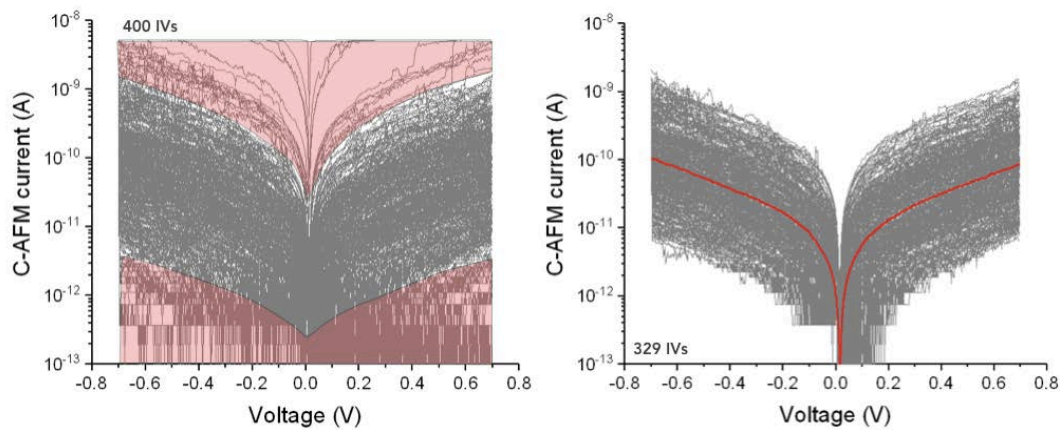

*Figure S7.* *Comparison of the complete set of I-V traces (400 IVs) measured on the TPT(A)-Au pristine sample. The light red areas show the IVs traces discarded (see text) from the analysis, leading to 329 useful IVs.*

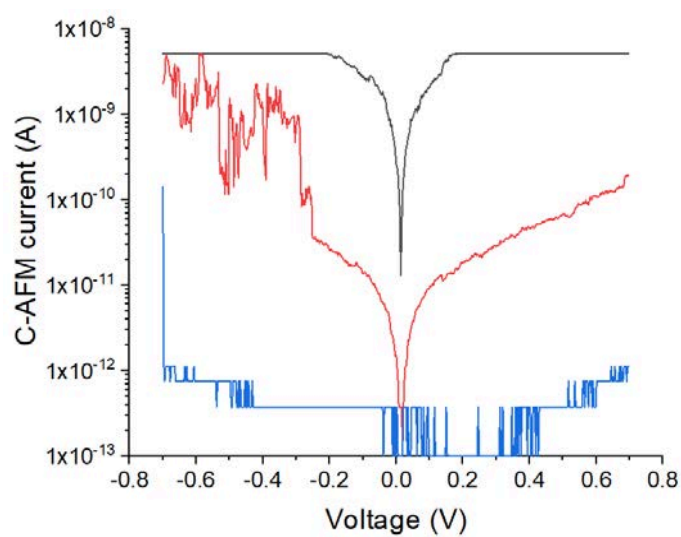

*Figure S8.* Typical examples of I-V curves discarded from the data analysis.

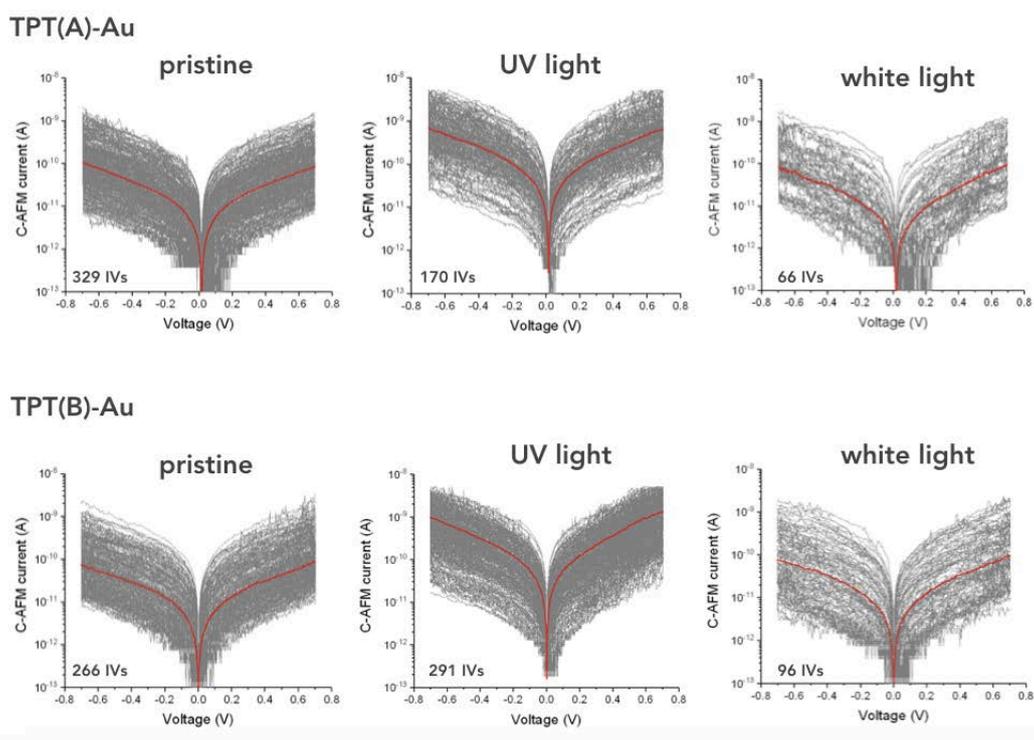

*Figure S9.* Datasets of the I-V measurements (semi-log scale) for the Au-TPT(A) and TPT(B) SAMs (pristine, after UV light illumination and visible light illumination). The red lines are the mean $\bar{I}$-V curves.

|  | Complete set | Analyzed set |
|---|---|---|
| Au-TPT(A) pristine/UV/vis | 400/400/400 | 329/170/66 |
| Au-TPT(B) pristine/UV/vis | 400/400/400 | 266/291/96 |
| Co-TPT(A) pristine/UV/vis | 400/400/400 | 141/364/185 |
| Co-TPT(B) pristine/UV/vis | 625/1250/625 | 225/514/107 |

*Table S2. Measurement yield.*

**Voltage dependent $R_{c/o}$.**

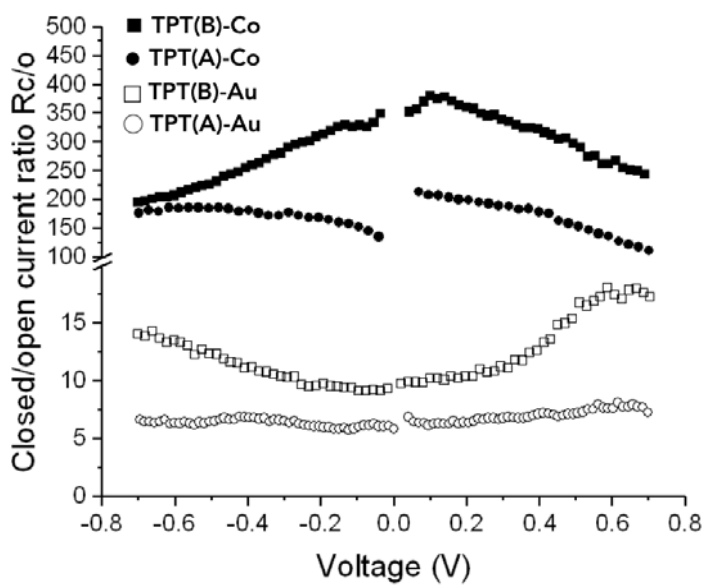

*Figure S10. Closed/open current ratio versus voltage $R_{c/o}(V)=\bar{I}_{UV}(V)/\bar{I}_{pristine}(V)$ for the four samples.*

**Fit of the energy level model.**

All the I-V traces in Figs. 3-4 (main text) were fitted individually with the single energy-level (SEL) model (Eq. 1, main text) with 3 fit parameters: $\varepsilon_0$ the energy position (with respect to the Fermi energy of electrodes) of the molecular orbital involved in the electron transport, $\Gamma_1$ and $\Gamma_2$ the coupling energy between the molecules and the two electrodes. The fits were done with the routine included in ORIGIN software, using the method of least squares and the Levenberg Marquardt iteration algorithm.

The SEL model is a low temperature approximation albeit it can be used at room temperature for voltages below the resonant transport conditions[29, 30] since the temperature broadening of the Fermi function is not taken into account. Moreover, a possible voltage dependance of $\varepsilon_0$ is also neglected.[31] It is known that the value of $\varepsilon_0$ given by the fit of the SEL model depends on the voltage window used for the fit.[29-31] This feature is confirmed (Fig. S10) showing that unreliable values are obtained with a too low voltage range (i.e. the SEL model is not reliable in the linear regime of the I-V curves) and not applicable when the voltage is high enough to bring the electrode Fermi energy close to molecular orbital (near resonant transport), here for a voltage window -0.7/0.7 V where all the values of $\varepsilon_0$ collapse. In the voltage windows -0.3/0.3 V to -0.6/0.6V we clearly observe a lowering of $\varepsilon_0$ upon UV illumination by around 0.1 eV for TPT(A) and 0.13 eV for TPT(B) - on average, a behavior also confirmed by the TVS (transient voltage spectroscopy) method (*vide infra*, Fig. S11).[32-37] For these reasons we limited the fits to a voltage window -0.5 V to 0.5 V to analyze the complete datasets shown in Figs. 3 - 4 (main text). To construct the histograms of the values of $\varepsilon_0$, $\Gamma_1$ and $\Gamma_2$ (Figs. 6 and 7), we discarded the cases for which the fits were not converging of not accurate enough (i.e. R-squared < 0.95). Typical fits on the mean Ī-V curves are shown in Fig. S11 for the two samples on $^{TS}$Au and the three conditions (pristine, after UV light, after white light).

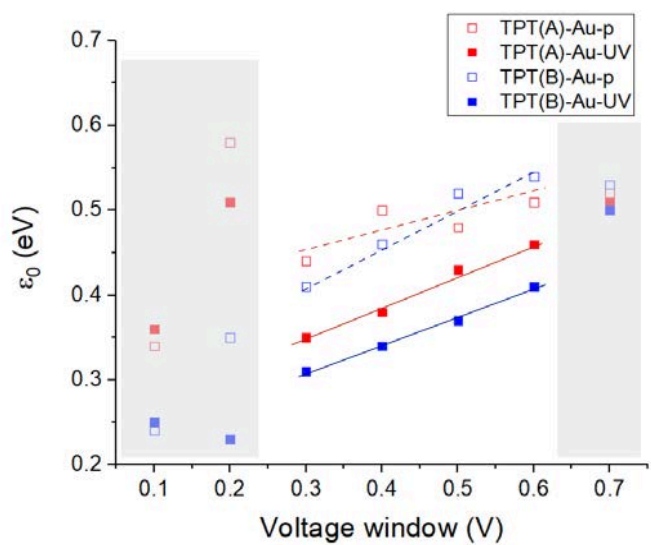

***Figure S11.*** *Values of $\varepsilon_0$ obtained with SEL model fitted on the mean $\bar{I}$-V curves for the two molecules on $^{TS}$Au (pristine and after UV illumination) with increasing voltage windows (-0.1/0.1 V to -0.7/0.7 V) for the fits.*

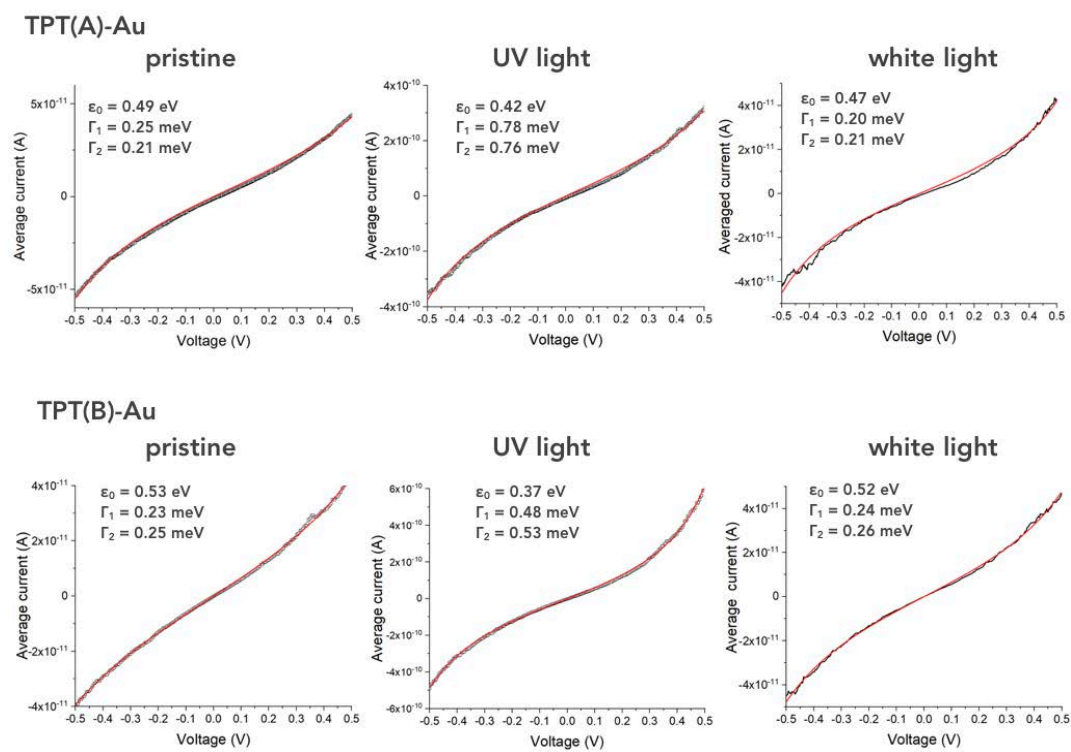

***Figure S12.*** *One energy level model fits on the mean current-voltage curves.*

The same mean Ī-V curves are also analysed by TVS, plotting $V^2/I$ (in absolute value) versus V (Fig. S12),[38] and determining the transition voltages ($V_{T+}$ and $V_{T-}$) for both voltage polarities, i.e. the voltage at the maximum of $V^2/I$. This threshold voltage indicates the transition between off-resonant (below $V_T$) and resonant (above $V_T$) transport regime in the molecular junctions. The values of $\varepsilon_0$ are estimated by:[36]

$$|\varepsilon_0| = 2\frac{e|V_{T+}V_{T-}|}{\sqrt{V_{T+}^2 + 10|V_{T+}V_{T-}|/3 + V_{T-}^2}} \quad (S4)$$

and they are marked in Fig. S12. They are in good agreement with the SEL fits (Fig. S11).

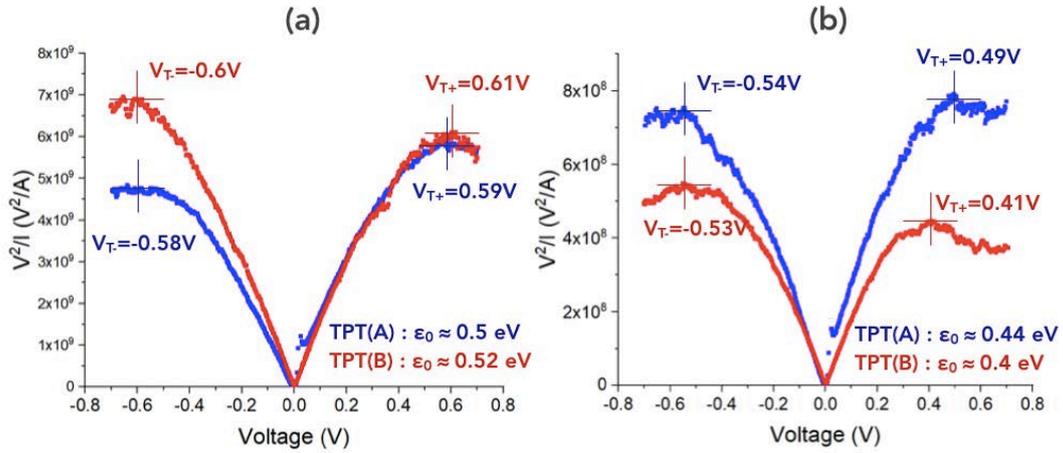

*Figure S13.* Typical TVS plots (|$V^2/I$|) vs. V. (a) TPT(A)-Au (blue) and TPT(B)-Au (red) in the pristine sate (open), (b) same samples after UV illumination (closed state). The thresholds $V_{T+}$ and $V_{T-}$ are indicated by the cross (with value) as well as the estimated values of $\varepsilon_0$ using Eq. (S4).

### Section 7. Illumination setup.

We used a power LED (M365FP1 from Thorlabs) for UV light irradiation (CAFM in air). This LED has a wavelength centered at 365 nm (close to the absorbance peak, see Fig. S1) and a bandwidth of 8 nm. An optical fiber was brought close (ca. 1 cm) to the sample in the CAFM setup (optical power density at the sample location ca. 27 mW/cm²). A chromatographic UV lamp (Vilbert-Lourmat, with a sharp peak at 365 nm and a background centered at 350 nm, BWHM: ~330-370 nm) was used for the measurements with the UHV CAFM and the irradiation of the sample was performed in the entry lock (P = 10⁻⁶ mbar $N_2$) of the instrument (optical power density at the sample location, ca. 10 cm is ca. 0.65 mW/cm²). For the visible light irradiation, we used a white light halogen lamp (Leica CLS150X) with a bandwidth centered at 600 nm (BWHM: ~500-700 nm), matching the absorbance peak of the closed form of TPT (Fig. S1) (optical power density at the sample location: ca. 220 mW/cm² in air at ca. 1 cm and ca. 13 mW/cm² at 10 cm for the experiments in UHV). Under these conditions, the samples were exposed to light for 1-3 h in air and 10-15 h in UHV, corresponding to almost the same photon density received by the sample, typically ~ 10²⁰ photons/cm². These conditions correspond to photostationnary states and we did not observe significant CAFM current variations with longer duration of light exposure.

## Section 8. Theoretical methods and additional calculations.

### *Simulated I-V curves.*

The I-V characteristics have been calculated on the basis of the Landauer-Büttiker formalism, which links the transmission spectrum to the current in a coherent transport regime.[39] When a bias is applied, the current is calculated via the integration of the transmission spectrum within a bias window defined by a Fermi-Dirac statistics in the left and right electrodes:

$$I(V) = \frac{2e}{h} \int T(E) \left[ f\left(\frac{E-\mu_R}{k_B T_R}\right) - f\left(\frac{E-\mu_L}{k_B T_L}\right) \right] dE \quad \text{(S5)}$$

where T(E) is the transmission spectrum, E the incident electron energy, f the Fermi function, $\mu_{R/L}$ the chemical potential of the right/left electrode, $T_{R/L}$ the temperature of the right/left electrode set here to 300K, $k_B$ the Boltzmann constant, e the elementary charge, h the Planck constant and V the applied bias. It is important to note that for an accurate estimation of the current, the transmission spectrum T(E) should be calculated in a self-consistent way for each bias. Thus, the current-voltage properties and the $R_{c/o}$ of the Au-TPT/Au junctions were predicted by using the transmission calculated at each bias, which is not too probihitive at the computational level. However, it is possible to obtain a reasonable approximation for the current at low bias by using the transmission spectrum at zero bias. This approximation is required for Co-TPT/Au junctions with a large unit cell and a spin-polarized electrode because the self-consistent calculations become very time consuming. Accordingly, the current-voltage properties of Co-PTP/Au and the corresponding $R_{c/o}$ were predicted by using the transmission calculated at zero bias.

*Lorentzian fitting: Γ broadening of Au-TPT(A)/Au junction transmission peaks.*

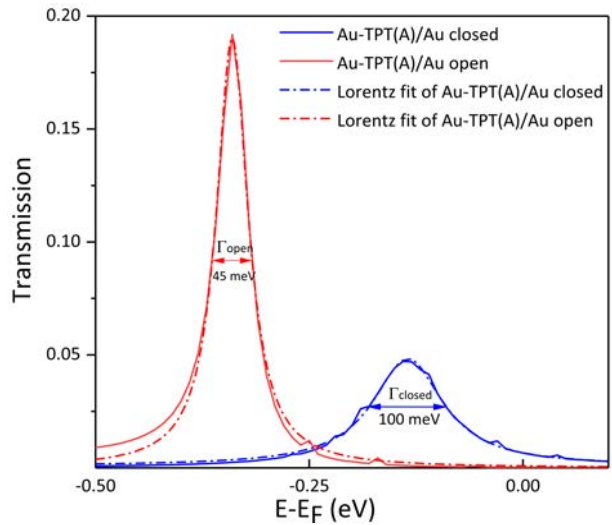

*Figure S14. Lorentzian fitting of the transmission peak of Au-TPT(A)/Au junction in both closed and open forms. The fitted Γ marked by an arrow indicates that the closed form exhibits larger broadening (100 meV) compared to open form (45 meV).*

***Au-TPT/Au molecular junctions : non tilted configuration.***

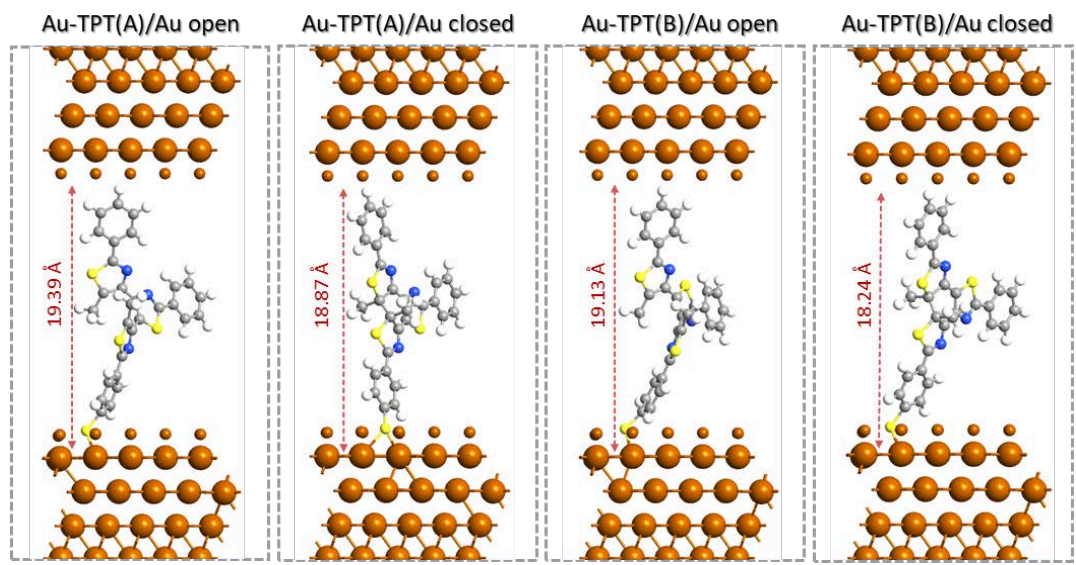

***Figure S15.*** *Optimized Au-TPT/Au junctions in a non tilted configuration. The calculated junction thickness is also marked. The small brown atoms refer to the gold ghost atoms.*

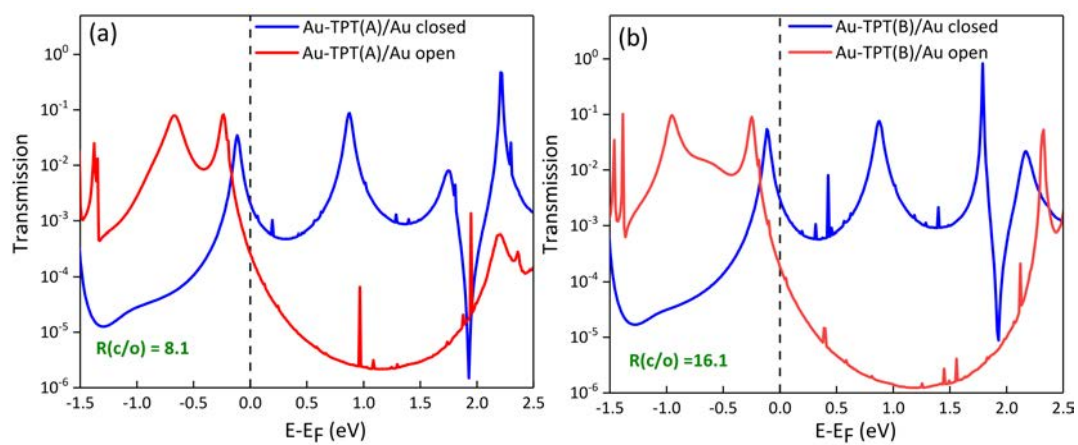

***Figure S16****. Log scale plot of the transmission spectra at zero bias for non tilted (a) Au-TPT(A)/Au and (b) Au-TPT(B)/Au junctions in their closed and open forms. The calculated $R_{c/o}$ are 8.1 and 16.1, respectively.*

## HOMO evolution as a function of the bias for Au-TPT/Au junctions.

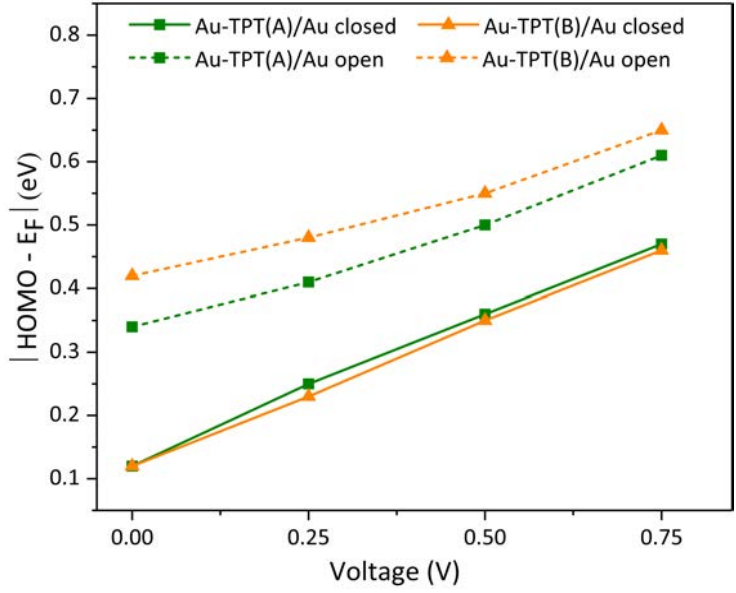

**Figure S17.** *The HOMO level evolution with respect to the average Fermi level of the electrodes as a function of the bias.*

## Charge transfer at the interface.

The charge rearrangement upon bond formation between the metal surface and the molecule[40] is defined as the difference between the plane averaged charge density of the full metal-SAM system, $\rho_{sys}$ and the sum of the density of the isolated subsystems, the free metal surface $\rho_{slab}$ and the free-standing molecules $\rho_{SAM}$:

$$\Delta \rho(z) = \rho_{sys} - \rho_{slab} - \rho_{SAM} \tag{S6}$$

For a deeper understanding of the implications of the charge rearrangements at the metal-SAM interface, we calculate the net charge transfer at the interface (ΔQ) by integrating the charge density redistribution (Δρ) along the z normal axis.

$$\Delta Q(z) = \int_0^z \Delta \rho(z) dz \tag{S7}$$

This gives the total amount of charge transferred from the left to the right of a plane lying at the position z. Here, the electronic density of the free metal surface (the isolated molecule) is calculated by removing the molecule (metal surface) from the functionalized system while keeping the same geometry as in the full system. Note that we describe here the chemisorption process in a radical scenario depicting the formation of a covalent bond between the molecule in its radical form and the metal surface.[40-42]

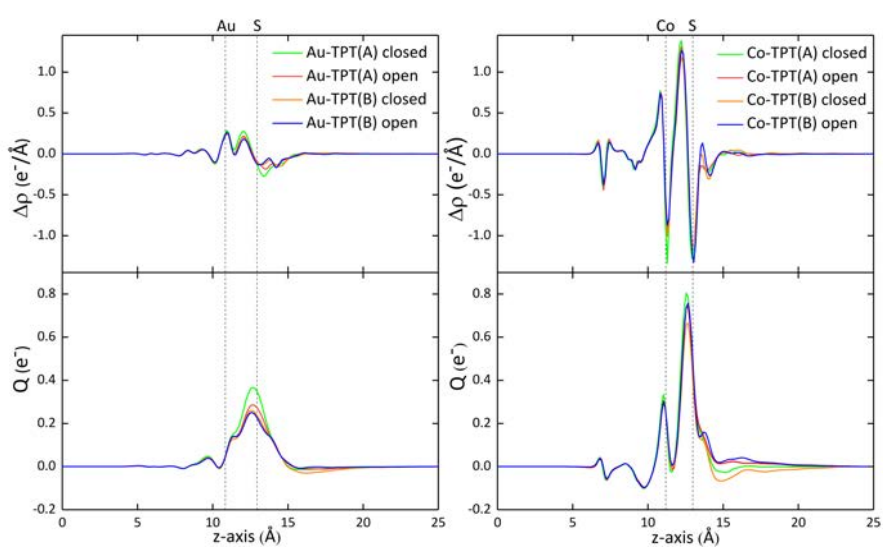

*Figure S18. Plane averaged charge density (top) and cumulative charge transfer (bottom) along the normal axis to the metal surface for Au-TPT (left) and Co-TPT(right). The dashed straight vertical lines represent the position of the first Au (Co) layer and S anchoring atom.*

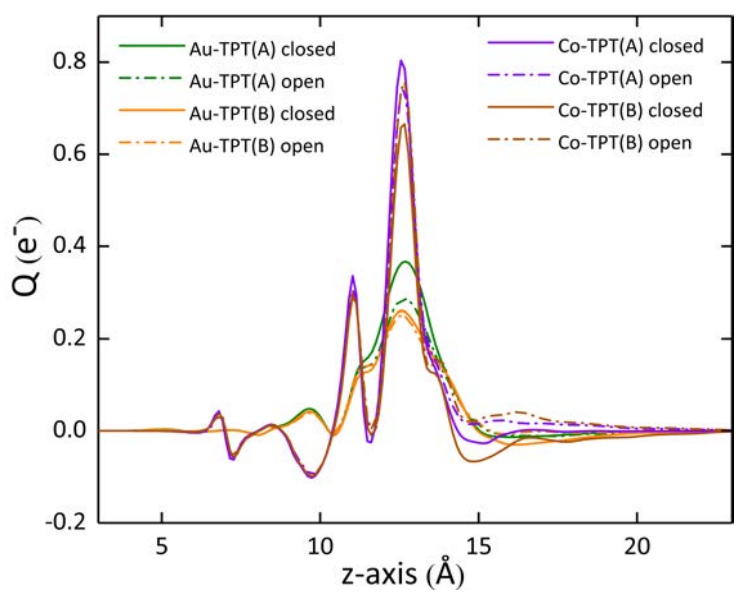

***Figure S19***. *Cumulative charge transfer along the normal axis to the metal surface for Au-TPT versus Co-TPT SAMs. The net charge transfer between the metal surface and TPT molecules is significantly larger for Co-TPT SAMs compared to Au-TPT SAMs.*

**Spin-dependent transmission spectra.**

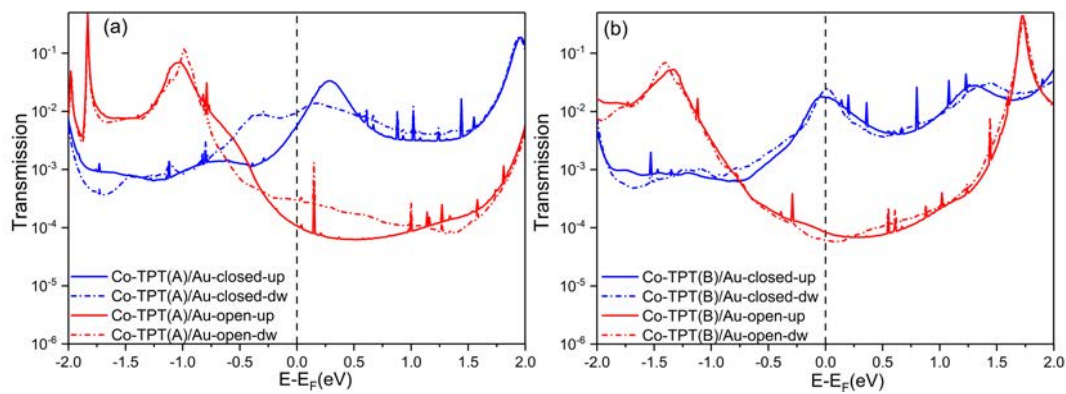

***Figure S20.*** *Log scale plot of the spin up (solid line) and spin down (dashed line) transmission spectra at zero bias for (a) Co-TPT(A)/Au and (b) Co-TPT(B)/Au junctions in their closed and open forms.*

***Zero-bias vs. finite bias voltage transmission spectra.***

We have calculated the closed/open ratio for Au-TPT/Au systems on the basis of the zero transmission spectra (see Table S3 below).

| Bias (V) | Au-TPT(A)/Au | | Au-TPT(B)/Au | |
|---|---|---|---|---|
| | Finite-bias | Zero transmission | Finite-bias | Zero transmission |
| 0 | 26.5 | 26.5 | 69.1 | 69.1 |
| 0.25 | 19.7 | 28.0 | 63.8 | 115 |
| 0.5 | 20 | 6.6 | 52 | 54 |
| 0.75 | 43.8 | 0.8 | 86.2 | 5 |

***Table S3.*** *Closed/open ratios ($R_{c/o}$) for the Au-TPT(A)/Au and Au-TPT(B)/Au junctions calculated using voltage-dependent transmission spectra (finite-bias) versus the zero-bias transmission spectra.*

By using the zero transmission spectrum, we obtain the same trend as with the finite-bias calculations: the Au-TPT(B) exhibits higher closed/open ratio compared to Au-TPT(A). However, the discrepancy associated t the use of the zero transmission spectra for estimating the closed/open ratio magnitude is sensitive to the voltage and the studied system (TPT(A) or TPT(B)), with a reasonable agreement found at 0.25V and 0.5V between the two options. We could then conclude it is reasonable to use the zero-bias transmission to compare with experimental results measured at 0.5V.

However, we consider these results obtained for a gold substrate are not directly transferrable to cobalt substrates. In fact, the TPT molecules exhibit a stronger coupling to cobalt that could result in a very different voltage drop. In other words, the magnitude of the discrepancy between the zero and the non-equilibrium transmissions highly depends on the investigated junction.

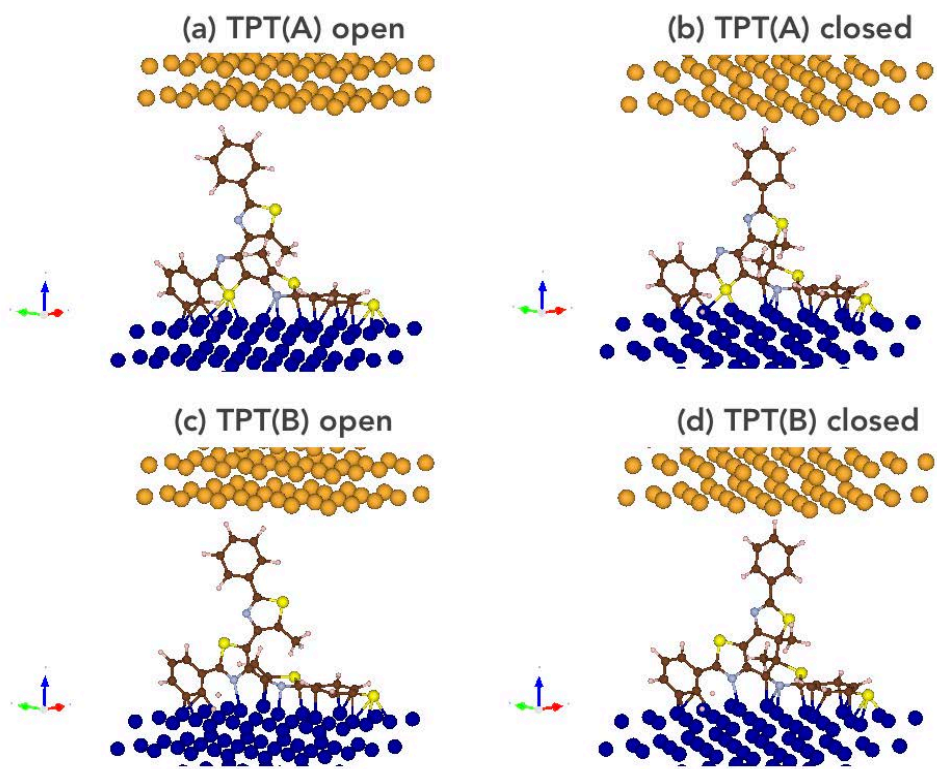

*Figure S21.* Closer view of Co-TPT interfaces: (a) and (b) TPT(A) molecule with only one N atom of the thiazole interaction with the Co surface, (c) and (d) TPT(B) molecule with 2 N atoms of the thiazole units interacting with the Co surface.

***Co-TPT/Au molecular junctions: non tilted configuration.***

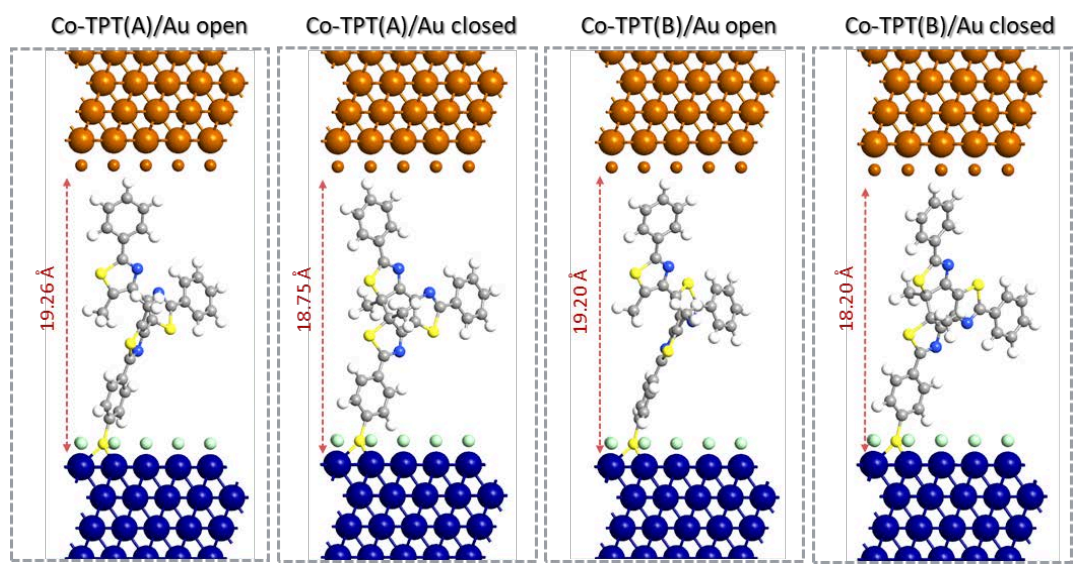

***Figure S22***. *Optimized non tilted Co-TPT/Au junctions. The calculated junction thickness is also marked. The small brown (green) atoms refer to gold (platinum) ghost atoms.*

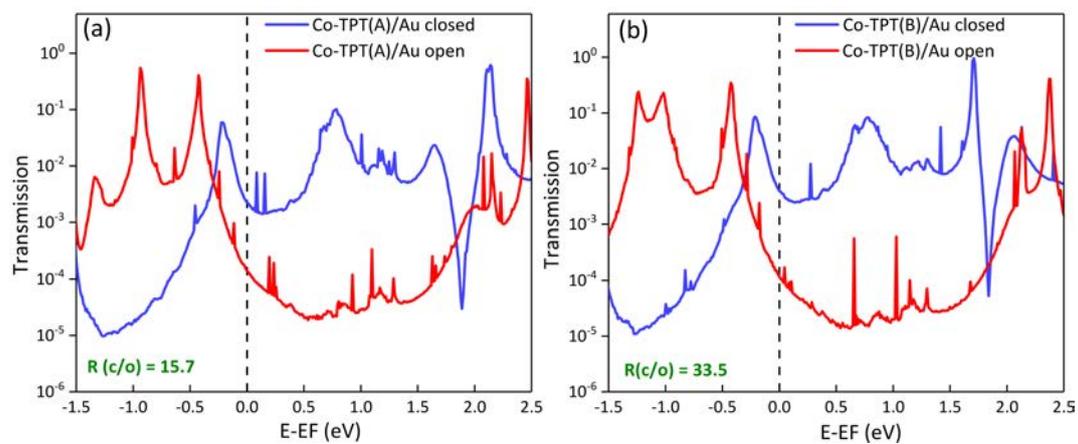

***Figure S23***. *Log scale plot of the transmission spectra at zero bias for non tilted (a) Co-TPT(A)/Au and (b) Co-TPT(B)/Au junctions in their closed and open forms. The calculated $R_{c/o}$ are 15.7 and 33.5, respectively.*

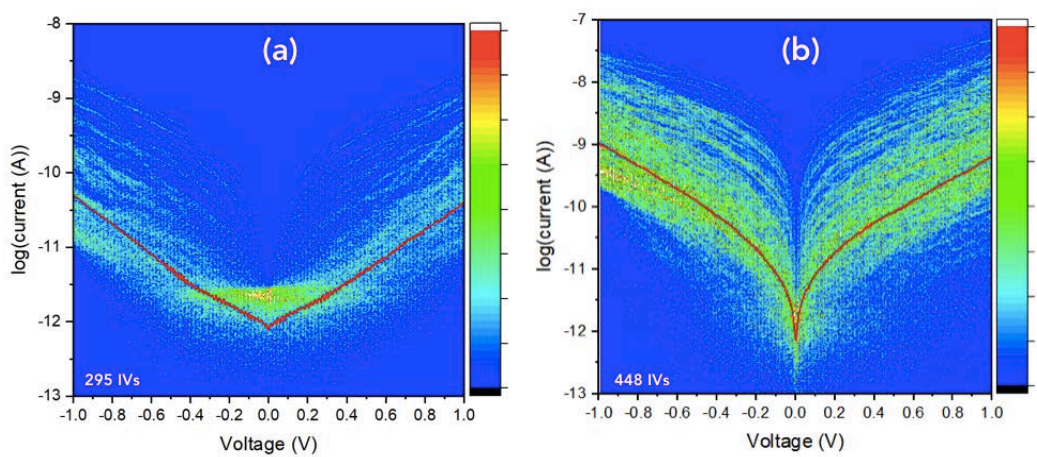

***Figure S24.*** *2D histograms of the current-voltage (I-V) curves: (a) pristine SAM of TPT(B) on Co (batch #2), (b) after UV irradiation. The currents are measured by CAFM in UHV. The number of I-V traces in the dataset are shown on the figures. The red line is the mean Ī current. From the mean current, the ratio $R_{c/o}$ is 15-25.*